# Heating in current carrying molecular junctions

Dvira Segal and Abraham Nitzan

School of Chemistry, The Sackler Faculty of Science, Tel Aviv University,
Tel Aviv, 69978, Israel

## Abstract

A framework for estimating heating and expected temperature rise in current carrying molecular junctions is described. Our approach is based on applying the Redfield approximation to a tight binding model for the molecular bridge supplemented by coupling to a phonon bath. This model, used previously to study thermal relaxation effects on electron transfer and conduction in molecular junctions, is extended and used to evaluate the fraction of available energy, i.e. of the potential drop, that is released as heat on the molecular bridge. Classical heat conduction theory is then applied to estimate the expected temperature rise. For a reasonable choice of molecular parameters and for junctions carrying currents in the nA range, we find the temperature rise to be a modest few degrees. It is argued, however, that using classical theory to describe heat transport away from the junction may underestimate the heating effect.

# 1. Introduction

In several recent publications we[1-4] and others[5-13] have addressed the issue of thermal effects in charge transport through molecular nano-junctions. There are two important reasons for the interest in this issue. First, on the fundamental level, the effect of electron-phonon[14] coupling is an important factor affecting the nature of the transmission and the conduction properties of the molecular junction. Secondly, from the practical point of view, this coupling is associated with possible heating of the junction as it operates as a conductor. As envisioned, among the most important advantages of molecular junctions is the combination of small size with versatile and controlled structure. On the other hand this small size implies small heat capacity, and possible heating may undermine the junction's structural integrity. This makes the understanding of heating effects in molecular conductors a crucial issue.

In this paper we study this issue using a simple model that combines an electronic system comprised of two continuous manifolds of states that represent the metal leads, a tight binding chain representing a molecular bridge that connects between these leads and a thermal phonon bath that couples to the molecular bridge. This model is similar to those used by us earlier[2-4] to study the effect of coupling to a phonon bath on the nature of the conduction process, in particular interplay between tunneling, activation and hopping transmission processes. Lake and Datta[15,16] have used a different approach based on the non-equilibrium Green's function formalism to study heat release in junction characterized by simple barrier or double barrier structures.

When a classical ohmic conductor characterized by a resistance $R$ carries a current $I$ the heat produced per unit time is $RI^2$. This translates into

$$W = \frac{J^2}{\sigma} \qquad (1)$$

where $W$ is the heat produced per unit time and volume, $J$ is the current density and $\sigma$ is the conductivity. In contrast, the observation of molecular scale resistance does not necessarily imply that heat dissipates locally on the source of this resistance. Consider for example a classical barrier separating between two identical reservoirs of charge carriers that are characterized by electrochemical potentials $\mu_L = \mu_R$ (Fig. 1a). Imposing a potential bias on this junction leads to the situation depicted in Fig. 1b, in which a steady state current flows in a closed circuit. This current is proportional to the rate difference

$$I = I_{L \to R} - I_{R \to L} = qA\left(e^{-\beta(E_B - \mu_L)} - e^{-\beta(E_B - \mu_R)}\right) \; ; \; \mu_R - \mu_L = q\phi \qquad (2)$$

where $A$ is the pre-exponential of the barrier crossing rate, $\Phi$ is the potential bias and $q$ is the carrier charge. For $q\phi \ll k_B T$ Eq. (2) yields $I = A\beta q^2 \phi$, which implies

$$G \equiv R^{-1} = A\beta q^2 \qquad (3)$$

Thus the potential barrier is associated with a resistance in this classical transport process, however the heat dissipation given by Eq. (1) should be considered more carefully. First, the net power $I\phi$ dissipated during this process is only a small fraction of the energy accumulated and then released as each carrier traverses the barrier. Secondly, this net dissipation does not necessarily fall on the barrier. In fact, in the common case where $A$ is derived from transition state theory, friction is assumed to play a negligible role on the barrier and the power $I\phi$ is dissipated in the side reservoirs rather than on the barrier. A similar phenomenon occurs in tunneling junctions where the Landauer conductance[17,18]

$$G(E_F) = \frac{e^2}{\pi\hbar}\mathcal{T}(E_F) \qquad (4)$$

($e$ is the electron charge, $\mathcal{T}$ is the transmission coefficient and $E_F$ is the Fermi energy) arises from elastic transmission and is not associated with any dissipation in the barrier. In both cases the net power $I\phi$ is dissipated in the leads, far from the barrier that represents the molecular junction.

For practical issues regarding heating effects on junction stability, the question where and how much heat is being released during conduction is of utmost importance. Energy dissipated as heat in the metal leads is expected to move away from the molecular junction relatively rapidly. On the other hand, energy released on the molecular bridge can potentially cause a large temperature increase due to the combination of relatively inefficient heat conduction away from the molecule with a relatively small heat capacity of the molecule itself. The Landauer formula (4) corresponds to the limit where dissipation of electronic energy on the barrier is absent, while dissipation in the metal is admitted only implicitly as discussed above. In reality, the coupling of electronic and nuclear degrees of freedom provides a mechanism for heat dissipation on the bridge itself. In the present paper we provide a framework for discussing this issue and for estimating the expected temperature rise on the bridge.

A crucial element of any analysis concerning heat release on the bridge is the distribution of the electrostatic potential drop on it. This issue has been discussed recently by several workers,[19-22] however no firm conclusions exist for any realistic system. Figure 2 shows several possible scenarios of the potential profiles between leads 1 and 2 when the potential bias is $\mu_1-\mu_2=\phi$. The linear ramp, $A_1A_2$, represents a commonly made assumption for metal-molecule-metal junctions with a strong chemical bonding of the molecule to both metals. Alternatively, in a scanning tunneling microscope (STM) experiment, a common assumption is that the electrostatic potential on the molecule is pinned to that of the substrate (lead 1 say) so that the entire potential drop occurs between the molecule and the tip (lead 2), leading to profile $A_1CA_2$. Because the molecule is a polarizable object we expect that the linear ramp potential should be replaced by the dashed line in the figure, that is sometimes approximated by the profile $A_1B_1B_2A_2$.[20,22]

A typical molecular junction carrying a current of 1nA through a potential drop of 0.5V, say, can deposit a power of up to $3 \cdot 10^9$ eV/s into the junction region. Such magnitude of heat power dissipated on a molecular bridge would pose a serious problem with regard to the bridge's structural integrity. The discussion above implies that $I\phi$ is only an upper bound, and that only a fraction, $\alpha I\phi$ $(\alpha < 1)$, is dissipated on the bridge itself. Estimating $\alpha$ is thus a central issue of our study.

In Section 2 we introduce our model and notations. Section 3 discusses a classical version of our problem where the molecular bridge is represented by a potential barrier separating two reservoirs of classical independent charge carriers (Fig. 1) that move under the influence of stochastic noise and damping. Section 4 discusses the quantum problem introduced in Sect. 2, using for the molecular bridge a tight binding model supplemented by a thermal bath and by a system-bath coupling. This model has all the ingredients of the classical model and also involves issues of coherence, dephasing and tunneling that are missing in the classical analog. In section 5 we discuss local aspects the heating process and provide an approximate method to compute the heat released at any local site of the bridge. In Section 6 we estimate the temperature rise on the molecular bridge under typical operating conditions, using a classical model for the heat conduction away from the bridge. Section 7 concludes.

## 2. Model and Notation

We use the same model that was used before to analyze the thermal effects in electron transmission through molecular bridges. This model (Fig. 3) consists of a molecular bridge ($M$), two metal leads ($J=L,R$ for the left and right lead, respectively), a thermal bath ($B$) and interactions between bridge and leads and bridge and thermal bath. For details see Section 2 of Ref. 4). The bridge is described by a tight binding model with $N$ sites and one state localized on each site. These states will be numbered by $n=1,...N$ and taken for simplicity to be mutually orthogonal with nearest-neighbor couplings. The left and right metal leads $J=L,R$ are represented by continuous manifolds of states, $\{j\}=\{l\}, \{r\}$. The corresponding Hamiltonian is

$$H = H_M + H_B + F + H_J + H_{JM} \tag{5}$$

where $H_B$ is the Hamiltonian for the thermal environment and where

$$H_M = H_0 + V$$
$$H_0 = \sum_{n=1}^{N} E_n |n><n| \quad ; \quad V = \sum_{n=1}^{N-1} V_{n,n+1} |n><n+1| + V_{n+1,n} |n+1><n| \tag{6}$$

$$H_J = \sum_l E_l |l><l| + \sum_r E_r |r><r| \tag{7}$$

$$H_{JM} = \sum_l V_l + \sum_r V_r$$
$$V_l = V_{l,1} |l><1| + V_{1,l} |1><l| \qquad V_r = V_{r,N} |r><N| + V_{N,r} |N><r| \tag{8}$$

$$F = \sum_{n=1}^{N} F_{n,n} |n><n| \tag{9}$$

In the calculation presented below we consider a particular version of this model in which $V_{n,n+1} \equiv V$ are the same for all nearest neighbors, and also all bridge energies $E_n$ ($n=1,...,N$) are taken equal, $E_n = E_B$, in the unbiased case. This model is depicted in Fig. 3, which also shows a particular incident state $|0>$ of the left manifold with energy $E_0 = E_B - \Delta E$, as well as the coupling to the thermal bath B. This coupling is taken to be of the form (9), where again $\{n\}$ is the set of $N$ bridge states in the site representation and where $F_{n,n}$ are operators in the bath subspace. These operators are characterized by their time correlation functions

$$\int_{-\infty}^{\infty} dt e^{i\omega t} <F_{n,n}(t)F_{n',n'}(0)> = e^{\beta\hbar\omega} \int_{-\infty}^{\infty} dt e^{i\omega t} <F_{n',n'}(0)F_{n,n}(t)> \quad ; \quad \beta = (k_B T)^{-1} \quad (10)$$

where $T$ is the temperature and $k_B$ – the Boltzmann constant. For specificity we sometime use

$$<F_{n,n}(t)F_{n',n'}(0)> = \delta_{n,n'} \frac{\kappa}{2\tau_c} \exp(-|t|/\tau_c) \tag{11}$$

in which $\kappa$ and $\tau_c$ play the roles of coupling strength and correlation time, respectively. The RHS of Eq. (11) becomes $\kappa\delta(t)$ in the Markovian, $\tau_c \to 0$, limit.

Our model is then characterized by the bridge length $N$, the energy gap $\Delta E$, the intra-bridge coupling $V$, the bridge-leads coupling expressed be the damping rates $\Gamma$ and the thermal-coupling parameters $\kappa$ and $\tau_c$. Previous uses of such model have yielded reasonable fits to the performance of actual molecular junctions taking N of order 10, $\Delta E$ in the range of a few thousands wavenumbers, and $V$ and $\Gamma$ in the range 100-1000cm$^{-1}$. Some information on the thermal coupling parameters associated with any given molecular site can be obtained using the formal relationship to the site reorganization energy $E_R$, $\kappa = k_B T E_R \tau_c / \hbar$.[24] In the model calculations described below we have used for simplicity the Markovian limit, $k_B T \tau_c / \hbar \ll 1$, and have taken $\kappa$ in the range $\kappa$=0.1-0.01$E_R$. (Typical reorganization energies are in the range of ~0.5eV).

In the absence of thermal interactions this model leads[4] to the following expression for the differential (per unit of the final energy range) transmission for an incoming electron with energy $E_0$

$$\mathcal{T}'(E_0, E) = \delta(E-E_0)\mathcal{T}(E_0) = \delta(E-E_0) Tr_M \left( G^{(M)}(E_0) \Gamma^{(L)}(E_0) G^{(M)\dagger}(E_0) \Gamma^{(R)}(E_0) \right)$$
(12)

(We use $\mathcal{T}'$ to denote the differential transmission coefficient, while $\mathcal{T}$ is the (dimensionless) elastic transmission coefficient). In (12) $Tr_M$ is a trace over the subspace $\{n\}$ of molecular bridge states, and $G^{(M)}(E)$ is the Green's function associated with this subspace

$$G^{(M)}(E) = \left( E - \mathbf{H}^{(M)}(E) \right)^{-1} \tag{13}$$

$$H^{(M)}_{n,n'}(E) = E_n \delta_{n,n'} + V_{n,n'} + \Sigma_{n,n'}(E) \tag{14}$$

with $\Sigma$ being the self-energy associated with the interaction of the bridge states with the metal electrodes and $\Gamma$ – its imaginary part

$$\Sigma_{n,n'}(E) = \Sigma_{n,n'}^{(L)}(E) + \Sigma_{n,n'}^{(R)}(E)$$

$$\Sigma_{n,n'}^{(J)}(E) = \sum_j \frac{V_{n,j} V_{j,n'}}{E - E_j + i\eta/2} = \Lambda_{n,n'}^{(J)}(E) - \frac{1}{2} i \Gamma_{n,n'}^{(J)}(E) \ ; \quad J = L, R \quad (15)$$

The elastic transmission coefficient $\mathcal{T}(E)$ is related to the zero bias conduction of the junction by the Landauer formula, Eq. (4).

As mentioned in Sect. 1, the electrostatic potential profile along the biased junction is an essential element in our analysis. We do not determine this distribution in the present paper. Instead, we will consider two models that correspond to the situations depicted in Fig. 2. Using $\Delta E$ to denote $E_B$-$E_F$ in the unbiased junction, model A is defined so that $\mu_L = E_F + (1/2)e\phi$; $\mu_R = E_F - (1/2)e\phi$; $E_1 = E_F + \Delta E + (1/4)e\phi$; $E_N = E_F + \Delta E - (1/4)e\phi$ and a potential drop of $(1/2)e\phi$ is distributed linearly along the bridge between sites 1 and N, i.e. $E_n = E_{n-1} - e\phi/2(N-1)$ ; $n = 2,...,N-1$. In model B $\mu_L$, $\mu_R$, $E_1$ and $E_N$ are the same as in model A and the other bridge levels are taken independent of $\Phi$, $E_n = E_F + \Delta E$ ; $n = 2,...,N-1$. Fig. 4 shows schematic views of these two models.

## 3. Heat release – the classical analog

The following classical model contains the essential ingredients of our problem: The molecular bridge is represented by a potential barrier (Fig. 1), and the transmission is a classical process of barrier crossing. Any particle that traverses the barrier from left to right starts its trip on the barrier at $x=0$ and ends it as it leaves the barrier at the point $x=L$. The particles are assume independent and their motion is governed by the Newton equation supplemented by a Langevin white noise

$$\ddot{x} = -\gamma \dot{x} + \frac{1}{m} F + \frac{1}{m} R(t) \quad (16)$$

where $F = -dU(x)/dx$ is the force derived from the potential barrier and where the friction $\gamma$ and random force $R$ satisfy

$$< R(t) > = 0 \quad and \quad < R(t_1) R(t_2) > = 2\gamma m k_B T \delta(t_1 - t_2). \quad (17)$$

When the potential bias $\phi=(\mu_L-\mu_R)/(-e)$ is distributed uniformly over the barrier we have

$$U(x) = -e\phi \frac{x}{L} \ ; \qquad F = \frac{e\phi}{L} \quad (18)$$

We will consider this situation, which is the classical analog of Model A of Fig. 4. A classical treatment of transmission problem that analyzes the bridge length (L) dependence of the transmission probability was recently provided by Hershkovitz and Pollak.[23]

Note that by starting the particles on top of the barrier at $x=0$ (Fig. 1b) we disregard the energy needed to get there in the consideration of heat release along their descent. The question is simply what fraction $\alpha$ of the available potential energy $e\phi$ that the electron looses as it traverses the distance between $x=0$ and $x=L$ is dissipated as heat *on the barrier*. Obviously $\alpha \to 1$ as $L \to \infty$, but it will be smaller for $L$ of the order of, or smaller than the relaxation distance $L_r$ (the distance, of order $F/m\gamma^2$, beyond which the descending particle assumes constant velocity).

We will not dwell here on the full solution of this stochastic transport problem and will limit ourselves to the simple zero temperature case. Eq. (16) for this case, $\ddot{x} = -\gamma \dot{x} + F/m$, yields $v(t) = v_0 \exp(-\gamma t) + (F/m\gamma)(1-\exp(-\gamma t))$ and $x(t) = x_0 + v_0\left[(1-\exp(-\gamma t))/\gamma\right] + (F/m\gamma)\left\{t - \left[(1-\exp(-\gamma t))/\gamma\right]\right\}$, where $x_0=0$ and $v_0$ are the initial position and velocity. The time to reach the end of the slope is the solution $t^*$ of the equation

$$L = v_0\left[(1-\exp(-\gamma t^*))/\gamma\right] + (F/m\gamma)\left\{t^* - \left[(1-\exp(-\gamma t^*))/\gamma\right]\right\} \tag{19}$$

Taking for simplicity $v_0=0$, the fraction of energy dissipated into heat on the slope is obtained from

$$\alpha(L) = \frac{e\phi - (1/2)mv^2(t^*)}{e\phi} \tag{20}$$

In the limit $\gamma t \ll 1$ we find

$$t^* = \sqrt{2mL/F} = \sqrt{2mL^2/e\phi} \tag{21}$$

and the corresponding condition for this limit

$$\gamma\sqrt{2mL^2/e\phi} \ll 1. \tag{22}$$

When this condition is satisfied we find from (20) and (21)

$$\alpha(L) = \gamma L\sqrt{\frac{2m}{e\phi}} \tag{23}$$

In the opposite high-friction/long-conductor limit $v(t^*)$ assumes its saturation value $v(t^*) = F/(\gamma m)$ and Eq. (20) yields

$$\alpha(L) = 1 - \frac{e\phi}{2m\gamma^2 L^2} \tag{24}$$

We conclude that in the low friction limit (Eq. (22)) $\alpha(L)$ is proportional to $\gamma L$, while in the opposite limit $\alpha(L)$ approaches unity with a correction that vanishes like $(\gamma L)^{-2}$.

## 4. A quantum calculation of heat release

Model A (Fig. 4) depicts a version of our quantum mechanical model that is analogous to the classical system discussed above. In the quantum case the incoming state |0> pumps the system, leading to a final energy distribution characterized by a quasi-elastic tunneling component and a thermal component resulting from propagation on the bridge. These contributions are distinct from each other (see, e.g. Fig. 3 of Ref. 4) only when the incoming energy $E_0$ is well separated from the energy of the bridge levels. Such situations are not expected to be of concern with regard to heating problems, and we study them first as a matter of theoretical interest.

To be specific, consider the case where the incoming energy is considerably below the bridge levels. The thermal component in this case is the analog of the classical process discussed in Sect. 3. It can be envisioned as a process in which the electron starts on the level |1> with energy $E_1$ and is emitted into the right manifold with a lower average energy $<E>_T$. The difference $E_1 - <E>_T$ is the amount of heat released on the bridge. The fraction $\alpha$ of available energy that is released as heat on the bridge is then

$$\alpha = \lambda \frac{E_1 - <E>_T}{e\phi} \tag{25}$$

where $\lambda$ is the fraction of the flux that is transmitted by the thermally activated route and where $\phi$ is the potential drop, $e\phi = E_1 - E_N$. Note that as written, the numerator in Eq. (25) is the heat released on the bridge per transmitted electron. Again, the thermal energy needed to place the electron on the bridge, which is pumped out of the left lead, is not taken into account in the definition (25) of $\alpha$.

A framework for evaluating the energy distribution of a transmitted electron in a model exemplified by Figs. 3 and 4 has been described in Refs. 4 and 24.[25] For an incoming state of energy $E_0$ this calculation yields the thermal analog of the differential transmission coefficient $\mathcal{T}'(E_0, E)$, Eq. (12), which contains both elastic and inelastic contributions to the transmission. The average energy associated with the thermal flux is then given by

$$<E>_T = \frac{\int_T \mathcal{T}'(E_0, E) E dE}{\int_T \mathcal{T}'(E_0, E) dE} \tag{26}$$

while the factor $\lambda$ is given by

$$\lambda = \frac{\int_T \mathcal{T}'(E_0, E) dE}{\int_{-\infty}^{\infty} \mathcal{T}'(E_0, E) dE} \tag{27}$$

where $\int_T$ denotes an integral over the thermal part of $\mathcal{T}'(E_0, E)$. Obviously, these quantities can be defined only when the tunneling and the thermal component of the transmission flux are well separated on the final energy axis. Interestingly, we have found that in this case the factor $(E_1 - <E>_T)/(e\phi)$ of Eq. (25) depends only very weakly on the incoming energy $E_0$.

Results based on Eqs. (25)-(27) are shown in Figures 5-6. In these calculations the bridge and the bridge-leads couplings are characterized by the choice of parameters $V=200 cm^{-1}$ and $\Gamma_L=\Gamma_R=160 cm^{-1}$, where the other parameters are varied as indicated below. Fig. 5 shows the fraction $\alpha$ plotted against the voltage difference $\phi$ for $\Delta E = 3000 cm^{-1}$, $T=300K$, bridge lengths $N=5$ or $10$ and thermal coupling strengths $\kappa=50$ or $200 cm^{-1}$ (the thermal bath is assumed to be Markovian, $\tau_c=0$). Note that $\kappa$ here is the analog of the friction $\gamma$ used in Sect. 3. The fact that these quantities are proportional to each other can be seen from their relation to the diffusion constant: $D = (\beta m \gamma)^{-1}$ in the classical case of Sect. 3 and $D \sim l^2 k_{hop}$ with $l$ being the intersite distance on the bridge and $k_{hop}=4V^2/\kappa$.[2] Still, the behavior displayed in Fig. 5 shows an interesting difference from the classical results of Sect. 3 in that a minimum appears in the $\alpha(\Phi)$ curve. Such a minimum is not indicated by the limiting expressions (23) and (24), that show both a decrease in $\alpha$ with increasing $\phi$.[26] Furthermore, for the parameters used in Fig. 5 $\lambda$ is very close to 1, and the displayed dependence on $\phi$ is essentially a property of the factor $(E_1 - <E>_1)/e\phi$.

The correspondence to the classical model of Sect. 3 is seen also in Figure 6 where $\phi/(1-\alpha)$ is plotted against $\kappa^2$, with $T=800$K, $\Delta E=1000\text{cm}^{-1}$ and $\phi=200\text{cm}^{-1}$, for several bridge lengths $N$. As $\kappa$ increases this dependence becomes linear, in agreement with the classical high friction limit, Eq. (24). In the opposite, low friction limit we find that $\alpha$ depends linearly on $\kappa$ as seen also in Eq. (23), however a closer examination reveals that in the quantum case this linear behavior is dominated by $\lambda$, Eq. (27), which was already shown to depend linearly on $\kappa$ for small $\kappa.^2$

We emphasize again that, while the above discussion is of general interest as a problem in quantum transport, the limit considered is not very relevant to the problem of heating in current carrying molecular conductors. Next we turn to the more interesting case where the current is dominated by resonance transmission through the bridge, i.e. by injection energies close to the bridge levels. Here the elastic and thermal fluxes cannot be energetically distinguished, and the total current is given by[27]

$$I = \frac{e}{\pi\hbar}\int_{-\infty}^{\infty} dE_0 \int_{-\infty}^{\infty} dE \left[ \mathcal{T}'_{LR}(E_0,E,\phi) f(E_0)(1-f(E+e\phi)) \right. \\ \left. - \mathcal{T}'_{RL}(E_0,E,\phi) f(E_0+e\phi)(1-f(E)) \right] \quad (28)$$

In analogy, the heat left on the bridge per unit time is given by[30]

$$I_h = -\frac{1}{\pi\hbar}\int_{-\infty}^{\infty} dE_0 \int_{-\infty}^{\infty} dE \left[ \mathcal{T}'_{LR}(E_0,E,\phi) f(E_0)(1-f(E+e\phi)) \right. \\ \left. + \mathcal{T}'_{RL}(E_0,E,\phi) f(E_0+e\phi)(1-f(E)) \right] (E-E_0) \quad (29)$$

Here $f$ is the Fermi-Dirac function, $\mathcal{T}'_{LR}$ and $\mathcal{T}'_{RL}$ are the transmission coefficients in the left-to-right and right-to-left transmission and the dependence on the finite voltage drop $\Phi$ across the junction (that makes $\mathcal{T}'_{LR}$ and $\mathcal{T}'_{RL}$ potentially different from each other) was written explicitly. The differential transmission coefficients $\mathcal{T}'(E_0,E)$ were introduced in Ref. 4, and are the thermal analogs of (12). The heat released on the bridge per transmitted electron is now obtained from Eqs. (28) and (29)

$$w = I_h e / I \quad (30)$$

It is important to realize that the results (25) and (29)-(30) arise from different approaches to different physical situations and are not equivalent. The result (25) corresponds to a quantum treatment of the process that underlies the classical discussion of Section 3. In this case the process that gives rise to heat release on the bridge is activated, and the heat release itself is associated with the flux of particles that had reached the top of the barrier at energy $E_1$, and are rolling down on the slope associated with the potential bias. Such particles injected with energy $E_0$ (a fraction $\lambda$ of the total number transmitted) have to gain energy of the order $E_1$-$E_0$ in order to start this process, but this energy gain is not taken into account in the computed energy balance. In contrast, Eq. (29) is a simple balance between the incoming and outgoing particle energies. When applied to situations where the current is strongly activated, it will predict that the net heat release is negative in situations where the average energy of the incoming particles is lower than that of the outgoing particles (such situations may arise because transmitted particles must be thermally activated to enter the barrier). In appendix A we show that with suitable handling based on these considerations, the result (29) reduces to (25) in the limit of large gap between $E_0$ and the bridge levels. In appendix B we examine the dependence of $I_h$ on the potential bias $\phi$. *Specifically we show that $I_h(\phi)$ satisfies the obvious condition $I_h(0) = 0$ and, furthermore, that for small $\phi$ $I_h(\phi) \sim \phi^2$.*

We next consider some numerical examples based on Eqs. (28)-(30). The results shown below are obtained using the model of Fig. 3 with the parameters $\Delta E$=2000cm$^{-1}$, V=200cm$^{-1}$, $\Gamma_1^{(L)} = \Gamma_N^{(R)} = 160$cm$^{-1}$, $\kappa$=50cm$^{-1}$, $\tau_c$=0 and $T$=300K.

In Figure 7 the current $I$, Eq. (28), calculated for models A and B of Fig. 4 (see Sect. 2) for an N=4-site bridge, is displayed against the voltage drop $\phi$. Note that the structure of our model corresponds to transmission through either occupied or unoccupied levels of the bridge so only one side of the potential bias is considered. Including both electron and hole transmission in the model will not change the considerations involving heat release in any essential way. The calculated current-voltage characteristic shows marked sensitivity to the potential drop profile on the bridge as already discussed in Ref. 20. Fig. 8 shows both $I$ vs. $\phi$ and $I_h$ vs. $\phi$ for models A and B with $N$=4, and Fig. 9 shows $w$, Eq. (30), for both models, plotted against the applied bias. The ratio $w/e\phi$, which is a measure of the fraction of available energy that is released as heat on the bridge, the analog of $\alpha$ of Eq. (25), is shown in Fig. 10. Figures 11a,b display for models A and B the electron current and the ratio $w/e\phi$ as functions of the bridge length $N$ for two values of the applied voltage, below resonance $\phi$=0.1V and above it, $\phi$=0.5V. Figure 12 shows the ratio $w/e\phi$ as a function of the thermal coupling strength $\kappa$ for several choices of molecular parameters in models A and B.

The following observations can be made:

(1) Both the current, $I$, and the heat release rate, $I_h$, depend on the model used for the potential drop profile on the bridge, however both models yield similar orders of magnitude for these quantities. As intuitively expected, the heat release per electron is higher for model A that is characterized by a linear potential drop along the bridge.

(2) The fraction $w/e\phi$ of the available energy released as heat on the bridge, which is the analog of $\alpha$ of Eq. (25) and Section 3, increases as the transmission assumes increasing resonance character. For the parameters used in Fig. 10 we see a marked increase in this ratio as the voltage increases towards and beyond the resonance transmission threshold $\phi\sim0.3$V.

(3) This fraction also increases with increasing bridge length, and on general grounds is expected to approach unity for large N. Still, for moderate bridge lengths, $N \leq 10$, and for the (reasonable) parameters used in our calculation, only ~10% of the available energy is dissipated on the bridge. This translates to a heat release of the order 0.1eV per transmitted electron or ~ $10^9$eV per second for currents in the nA range.

(4) For resonance transmission the bridge length dependence of both $I$ and $I_h$ reflects specific properties associated with bridge levels going in and out of resonance with the injection energy range, on top of generic phenomenology discussed in our earlier work.[2,3] The oscillatory dependence on the bridge length $N$ seen in the dashed lines in Fig. 11 is a manifestation of the first issue: the transmission probability changes as bridge levels get in and out of resonance with the injection energy. Increasing the bridge length may bring more levels of the bridge into resonance, leading, at intermediate bridge length to a counter intuitive increase of conduction with $N$, as seen in the dashed and dotted line of Fig. 11a. At the same time, the difference between the $N$ dependence at $T$=300K and $T$=200K is associated with the fact that at room temperature and for the parameters used transmission is dominated by thermal activation into the bridge, while at the lower temperature and small voltage the $I/N$ dependence at small $N$ shows the exponential behavior typical to tunneling, which crosses over to an algebraic dependence for large $N$.[2,3]

(5) As stated above (see appendix B), at small $\phi$ Eqs. (28) and (29) yield the expected Ohmic behavior $I \sim \phi$ and $I_h \sim \phi^2$. This is an important check on our formalism because it is not immediately obvious that Eq. (29) indeed satisfies $I_h(\phi = 0) = (dI_h / d\phi)|_{\phi=0} = 0$. As a check we have verified that our numerical code also shows this behavior.

(6). The small initial drop seen in the crossed ($T$=200K, $\phi$=0.1V) curve of Fig. 11b reflects the fact that in this (tunneling dominated) regime the small thermal contribution (whose importance increases with bridge length) causes transmission of particles at energies higher than the injection energy (see discussion below Eq. (30)).

With reasonable model parameters and under reasonable operating conditions Figs 8-10 tell us to expect that a substantial amount of energy 10-30% of the potential drop, will be released as heat on the bridge (see e.g. Fig 12 and recall that a reasonable choice for $\kappa$ is in the range 20-200cm$^{-1}$ (see Sect. 2)). Where on the bridge is this heat released and what is the expected temperature rise are the next questions on our agenda.

## 5. Local aspects of heat release

In the previous section we have shown how the heat release rate associated with electron transmission through a molecular junction can be computed within a simple model for the bridge. It is also of interest to ask where on the bridge this heat is released. For a bridge uniformly made of identical repeat units and attached symmetrically to two identical electrodes one may expect that heat generation will be uniform along the bridge, at least far enough from the molecule-lead surface contacts. It is of interest to consider other situations, e.g. the heat generated about an impurity site on the bridge structure or at special bonds, e.g. that connecting the molecule to the electrode surface. In this section we consider this issue within the same tight binding bridge model used above.

Again we consider a steady state pumped by an incoming state |0> in the manifold that represents the left metal lead. Denote by $J_k(E)dE$ the steady state probability flux at bridge site $k$ in the energy range $E...E+dE$. The integrated flux,

$$J = \int dE J_k(E), \tag{31}$$

is obviously the same for all sites. The average energy of the transmitted flux at site $k$ is given by

$$<E>_k = \frac{\int dE\, E J_k(E)}{J} \quad (32)$$

Knowledge of $J_k(E)$ at every bridge site k therefore suffices for evaluating the local heat dissipation during electron transmission: the averaged energy released as heat between sites $k$ and $k+1$ is simply $<E>_k - <E>_{k+1}$.

A way to calculate $J_k(E)$ is provided by a generalization of the procedure[4,24] that yields $\mathcal{T}'(E_0, E)dE$, the final-energy resolved differential transmission probability into energy range $E...E+dE$ in the manifold that represents the right metal lead, for a given incident energy $E_0$. This generalization (Fig. 13) is done by attaching to each bridge state $k$, ($k=1,...,N-1$) a fictitious electrode represented by the continuous manifold $K$ in Fig. 13. Only state $k$ of the bridge is coupled to states in its associated manifold, and this coupling is taken to be vanishingly small so that the main flux through the bridge is not affected by it. The same procedure that yields the energy resolved flux $\mathcal{T}'(E_0, E)$ into the right metal lead, can be used to get the corresponding flux $\mathcal{T}'_k(E_0, E)$ into the manifold $K$. We will now *assume* that for a given incident energy $E_0$, $\int dE_0 f(E_0) \mathcal{T}'_k(E_0, E)$ and $J_k(E)$ represent, up to normalization factors, the same quantity, so that the normalized energy distribution at site $k$ is

$$P_k(E) = \frac{\int dE_0 f(E_0) \mathcal{T}'_k(E_0, E)}{\int dE_0 f(E_0) \int dE\, \mathcal{T}'_k(E_0, E)} \quad (33)$$

and the average electron energy on site $k$ is

$$<E>_k = \int dE\, E\, P_k(E) \quad (34)$$

It should be emphasized the validity of Eqs. (33) and (34) is an assumption. Keeping in mind that the transmission coefficient $\mathcal{T}'$ that appears in Eq. (33) corresponds to what was denoted $\mathcal{T}'_{LR}(E_0, E, \phi)$ in Eq. (28), the integrated flux $J$ is given by

$$J = \int_{-\infty}^{\infty} dE_0 \int_{-\infty}^{\infty} dE \left[ \mathcal{T}'_{LR}(E_0, E, \phi) f(E_0)(1 - f(E + e\phi)) \right] \quad (35)$$

and depends on the availability of unoccupied states in the accepting final manifold. We have no theory for the effect of this availability on the intermediate quantities $J_k(E)$. It is only if $f(E+e\phi)=0$ in the relevant final energy range that Eqs. (33) and (34) are rigorously justified.

As a demonstration of this approach we show in Fig. 14 the computed integrated heat release, i.e. the heat generated between sites 1 and $n$ as a function of the site index $n$ for a system described by model B with the parameters $N=10$, $\Delta E=(E_B-E_F)=2000\text{cm}^{-1}$, $V=200\text{cm}^{-1}$, $\Gamma_1^{(L)} = \Gamma_N^{(R)} = 160\text{cm}^{-1}$, $\tau_c=0$, $\kappa=50\text{cm}^{-1}$ and $T=300$K, under a potential bias $e\phi = 8000\text{cm}^{-1} (\simeq 1\text{eV})$. This bias brings $\mu_L$ into resonance with the adjacent bridge level $|1>$. Setting the energy scale so that the unbiased value of the Fermi energy is zero, we have under this bias $\mu_L=E_1=4000\text{cm}^{-1}$, $\mu_R=-4000\text{cm}^{-1}$ and $E_N=0$, while the energies of the other bridge levels remain $E_B$ as defined by model B. To simplify the calculation we limit it to an initial energy equal to $\mu_L$, i.e. we take $f(E_0) = \delta(E_0 - \mu_L) = \delta(E_0 - E_1)$ in Eq. (33). Fig. 14 shows results for this model, as well as for systems with one impurity site, where $E_5=E_B$ is replaced by $E_5=E_B\pm 1000\text{cm}^{-1}$. As expected, we see that energy release occurs predominantly at the regions near the lead-molecule contacts that carry the potential drop. The local heat release (the slope of the lines in Fig. 14) initially increases, then decreases as the electron traverses a local low energy impurity (a smaller opposite effect is seen near a high energy impurity) but, except when the impurity is placed near a bridge edge, there is no significant effect on the overall heat release i.e. the value of $E_1 - <E_{10}>$ for the 10-site model studied. It should be emphasized however that this calculation is done for a given potential bias of 1eV. We find that the current calculated from Eq. (28) is $I/e= 1.10 \cdot 10^8 \text{s}^{-1}$, $9.94 \cdot 10^7 \text{s}^{-1}$, and $9.72 \cdot 10^7 \text{s}^{-1}$ for the no impurity case and for the cases with $E_5 = E_B - 0.125\text{eV}$ and $E_5 = E_B + 0.125\text{eV}$, respectively. Thus the presence of either impurity does increase the apparent junction resistance ($I/\phi$) by ~15%.

## 6. Estimating the temperature rise

We now turn our attention to the temperature rise expected in a current carrying bridge molecular conductor. In making the following estimate we disregard energy that is deposited directly into the leads. This assumes that heat conduction in the metal lead is efficient and that energy reaching the leads dissipates quickly into the bulk metals. On the other hand, energy released on the molecular bridge can be transferred only by nuclear degrees of freedom, i.e. by the process known in other contexts as intramolecular vibrational energy relaxation (IVR). IVR as a model of energy transfer in a molecule connecting two metal leads is an interesting problem that has not been considered yet, although some related work on heat transport in mesoscopic junctions has been recently published.[13,31-39] In this paper we limit ourselves to a much simpler approach based on the classical heat conduction of organic solids. In this rough model we represent the bridge by a cylinder of length $L$ connecting two planes (the metal surfaces) on which the room temperature $T_\infty$ is given (Fig. 15). Again, this assumes that heat conduction on the metal leads is very efficient relative to that on the bridge. This cylinder is comprised of two concentric cylindrical regions. The inner cylinder of radius $R_1$ is the current carrying region, and we assume that heat is generated uniformly on this region at a rate $i_h \equiv I_h/(\pi R_1^2 L)$ per unit volume. (In general this heat generation may depend on the position along the cylinder axis $z$ in a way that depends on the bridge structure and the potential drop profile, but in the present estimate this is disregarded). The outer cylinder of radius $R_2$ represents in this model regions on the molecular bridge on which heat is not deposited. In a microscopic model energy flows into the region $0 < z < L; R_1 < \rho < R_2$ is caused by redistribution of molecular nuclear energy (intramolecular vibrational relaxation, IVR), but here we will assume that energy flow in the molecule ($0 < z < L; \rho < R_2$) is governed by classical heat conduction characterized by an assumed known thermal conductivity $\sigma_h$. The temperature equation is then

$$\sigma_h \nabla^2 T + i_h = c \frac{\partial T}{\partial t} \tag{36}$$

Where $c$ is the heat capacity per unit volume. The temperature profile at steady-state is determined by the Poisson equation

$$\nabla^2 T = -\frac{i_h}{\sigma_h} \tag{37}$$

that should be solved under the given boundary conditions. On the left and right boundaries we have $T(z=0)=T(z=\infty)=T_\infty$. For the heat flow in the $\rho$ direction we consider two situations that give lower and upper bounds on the temperature rise:

(a) The molecular bridge is immersed in a condensed environment so that heat can be conducted away in the direction perpendicular to the current flow. A lower bound on the temperature rise on the bridge may than be obtained by imposing $T(\rho=R_2)=T_\infty$. This amounts to the additional assumption that thermal conduction in the surrounding environment is very fast. If we assume in addition that $L \gg R_2$ so that heat is dissipated mostly in the direction normal to the bridge, and disregard the contribution of heat loss through the electrodes, this yields(Ref. 40, Chap. 2-3)

$$T(\rho = 0) = T_\infty + \frac{i_h R_1^2}{2\sigma_h} \ln\left(\frac{R_2}{R_1}\right) + \frac{i_h}{4\sigma_h} R_1^2 \tag{38}$$

Obviously when $R_2 \to \infty$ we can no longer disregard the heat flux in the parallel direction, and Eq. (38) is no longer valid.

(b) An upper bound on the temperature rise on the bridge is obtained for a model that disregards all heat dissipation in the $\rho$ direction, i.e. by considering a bridge suspended between the two metal leads in vacuum, and disregard all radiative heat losses. In this case we need to solve Eq. (37) with the Dirichlet boundary condition $T(z=0) = T(z=L) = T_\infty$ on the bridge-metal interfaces, and a Neumann boundary condition $\left(\partial T / \partial \rho\right)_{\rho=R_2} = 0$ on the outer cylinder surface.

The Poisson equation (37) was solved using a standard finite difference algorithm(see e.g. Ref. 40 Chap. 3). Fig. 16 shows results obtained from this calculation, using typical molecular parameters. In particular we note that $\sigma_h = 10^{-4}$ cal/(s·cm·K) is a typical value for the heat conductivity of condensed organic materials. The heat generation rate $I_h = 10^{10}$ eV/s is the order of magnitude expected in a junction carrying a current of 10nA. We see that the temperature in the molecule increases only in a modest way that should not be significant in most situations. Obviously, for larger values of $L$, and when no heat flow is possible in the normal ($\rho$) direction, the temperature at the molecular center will be higher (we get $T \simeq 450$K for $L=500$Å). While these results are gratifying from the point of view of molecular conductors design, the crude nature of our approximations should be kept in mind. In particular a careful evaluation of vibrational energy flow in molecular bridges is highly needed and should be the next stage in this study.

## 7. Conclusions

Heating in current carrying molecular junctions is controlled by the combination of at least two factors. First, the amount, per transmitted electron, of electronic energy directly deposited on the molecular bridge is of great importance. Second, the rate of heat conduction away from the molecular bridge will determine the ultimate temperature rise at the junction. In this paper we have presented a framework for discussing these issues and for estimating the amount of temperature rise expected in current carrying single molecule conductors. To discuss the first issue we have developed a formalism that makes it possible to estimate the power dissipated on the bridge. In the strong localization limit, where the electron (or hole) fully thermalizes on each bridge site and charge carrier propagation proceeds by site-to-site hopping, all available energy (i.e. the full potential drop) is deposited as heat on the bridge. In this limit the heat power deposited on the bridge is given by the ohmic expression $I_h = I\phi$. In the opposite limit, where electron-phonon interactions on the molecular bridge are disregarded, no heat is deposited directly on the bridge. In intermediate cases only a fraction of the available energy will be deposited on the bridge. This fraction is expected to be small for large inter-site electronic coupling, strong bridge-lead coupling, relatively weak electron-phonon interaction and short bridges. Indeed, for a reasonable range of molecular and relaxation parameters we have found that this fraction may be substantially smaller than 1, even down to order 0.1, but given that in a junction that carries 1nA under a bias of 1V the total energy dissipation rate is $\sim 10^{10}$eV/s, and that less than 10eV is sufficient to dissociate the molecular bridge, the issue of temperature rise cannot be disregarded. This observation makes it imperative to consider the second factor - the efficiency of heat conduction away from the junction. This issue was treated in the present paper within a classical heat conduction model. For a simple model that represents the molecular bridge as a cylinder characterized by heat conductivity typical to organic solids, we have found the temperature rise in molecular junctions to be in the tolerable few degrees range even under the extreme conditions where all the energy associated with the potential bias is assumed to be deposited (uniformly) on the bridge, and where heat is allowed to escape only through the molecule-lead contact. It should be emphasized, however, that our classical heat conduction model is a gross oversimplification that is expected to underestimate the temperature rise. Not only does the classical theory of heat conduction expected to fail in restricted-geometry systems (see, e.g. Ref. 39), the discrete spectrum of nuclear motions in suspended molecular bridges may render vibrational energy transfer relatively slow. A molecular level treatment of vibrational energy transmission is needed to get a more reliable estimate of temperature rise in current carrying molecular junctions.

Finally we note that, while our generic treatment provides a framework for analyzing heating in molecular junctions, in practical applications one should worry about possible energy accumulation in specific molecular bonds. In particular, since much of the potential drop is expected to occur at the molecule-lead contacts, the possibility of heating these particular locations that are critical to the junction stability should be considered. In the present paper we have developed the theoretical framework for computing approximately the position dependence of the dissipated power, and have shown that one can associate this dependence with the bridge structure. Again, molecular level treatment of vibrational motions in specific molecular junctions will be needed to assess this issue.

**Appendix A**

Consider Eq. (29) and let $E_0$ be much below the bridge levels. We focus on the case where the potential bias $\Phi$ is positive so that $\mu_R < \mu_L$ and consider only the current from left to right. Eq. (28) then takes the form

$$I = \frac{e}{\pi\hbar} \int_{-\infty}^{\infty} dE_0 \int_{-\infty}^{\infty} dE\, T'_{LR}(E_0, E, \phi) f(E_0)(1 - f(E + e\phi)) \tag{39}$$

We will also assume that $\phi$ is much smaller than the gap between the injection and bridge energies. In this case the differential transmission from left to right may be approximated as a sum of coherent-elastic and thermal components

$$T'_{LR}(E_0, E, \phi) = A(E_0)\delta(E_0 - E) + e^{-\beta(E_B - E_0)} B(E_0, E) \tag{40}$$

Where the $A$ and $B$ terms are the coherent/tunneling and the activated components of the transmission, respectively. Under the approximations made the function $B(E_0, E)$, viewed as a function of the final energy $E$, is peaked in an energy range substantially higher (in terms of $k_B T$) than $E_0$, therefore $f(E_B)$ and $f(E_B + e\Delta\phi)$ may be taken to vanish. In this case Eq. (39) may be written as a sum of a tunneling component and a thermal component

$$I = I_{tun} + I_{therm} \qquad (41)$$

where

$$I_{tun} = \frac{e}{\pi \hbar} \int_{-\infty}^{\infty} A(E_0) f(E_0)(1 - f(E_0 + e\Delta\phi)) dE_0$$

$$I_{therm} \cong \frac{e}{\pi \hbar} \int_{-\infty}^{\infty} dE_0 f(E_0) e^{-\beta(E_B - E_0)} \int_{-\infty}^{\infty} dE B(E_0, E) \qquad (42)$$

The result for the heat generation rate (29) under the same approximation is obtained by modifying the thermal component in Eq. (42)

$$I_h \cong -\frac{1}{\pi \hbar} \int_{-\infty}^{\infty} dE_0 f(E_0) e^{-\beta(E_B - E_0)} \int_{-\infty}^{\infty} dE B(E_0, E)(E - E_0 - E_B) \qquad (43)$$

Note that in the spirit of our discussion below Eq. (30), the energy balance is computed by comparing the final energy to the energy of the activated electron at energy $E_0 + E_B$. The heat released per thermally transmitted electron is obtained by inserting expressions (42) and (43) into Eq. (30). We further assume (as was verified numerically above) that, in the limit considered, the integrals over $E$ are practically independent of $E_0$. We therefore get

$$\frac{I_h}{I_{therm}} = -\frac{\int_{-\infty}^{\infty} dE B(E_0, E)(E - E_0 - E_B)}{\int_{-\infty}^{\infty} dE B(E_0, E)} = E_1 - <E>_1 \qquad (44)$$

where $E_1 = E_0 + E_B$ and where $<E>_1$ is the average energy of the thermally transmitted electron. This heat release per thermally transmitted electron is then multiplied by the fraction of electrons transmitted thermally

$$\lambda = \frac{I_{therm}}{I_{tun} + I_{therm}} \qquad (45)$$

and divided by the energy available for release, $-e\phi$, to give the result (25).

**Appendix B**

Here we examine the dependence of the heat generation rate, $I_h$, Eq. (29), on the potential bias $\phi$. First consider the zero bias case. In this equilibrium case $I_h(\phi = 0)$ must vanish because there should not be net heat dissipation on the bridge in this situation. We will show that Eq. (29), which in this limit becomes

$$I_h(\phi = 0) = -\frac{2}{\pi\hbar} \int_{-\infty}^{\infty} dE_0 f(E_0) \int_{-\infty}^{\infty} (1 - f(E)) \mathcal{T}'(E_0, E)(E - E_0) dE \qquad (46)$$

indeed satisfies this requirement. For simplicity we limit ourselves to a model where the density of states in the two continuous manifolds does not depend on energy. In this case the differential transmission probability $\mathcal{T}'(E_0, E)$ is symmetric in its arguments, i.e. $\mathcal{T}'(E_0, E) = \mathcal{T}'(E, E_0)$. Since we are dealing with an equilibrium situation we may further assume that $\mathcal{T}'(E_0, E)$ satisfies the detailed balance condition

$$\mathcal{T}'(E_0, E) = \begin{cases} A(E - E_0)e^{-\beta(E - E_0)} & ; \quad E > E_0 \\ A(E - E_0) & ; \quad E \leq E_0 \end{cases} \qquad (47)$$

with $A(x) = A(-x)$. We take the chemical potential $\mu$ to be the zero reference energy, so that $f(x) = (1 + \exp(\beta x))^{-1}$. Rewriting Eq. (46) in the form $I_h(\phi = 0) = -(2/\pi\hbar)^{-1} \int_{-\infty}^{\infty} dE_0 F(E_0)$ with

$$F(E_0) = f(E_0) \left[ \int_{-\infty}^{E_0} (1 - f(E)) A(E - E_0)(E - E_0) + \int_{E_0}^{\infty} (1 - f(E)) e^{-\beta(E - E_0)} A(E - E_0)(E - E_0) \right] \qquad (48)$$

we will show that $F(E_0) + F(-E_0) = 0$, thus proving that $I_h(\phi = 0) = 0$. To this end we use the equalities

$$f(E_0)(1 - f(E)) = \frac{1}{e^{\beta E_0} + 1} \frac{e^{\beta E}}{e^{\beta E} + 1}$$
$$f(-E_0)(1 - f(E)) = \frac{1}{e^{-\beta E_0} + 1} \frac{e^{\beta E}}{e^{\beta E} + 1} \qquad (49)$$

with Eq. (48) to get

$$F(E_0)+F(-E_0) = \int_{-\infty}^{E_0} \frac{e^{\beta E}A(E-E_0)}{(e^{\beta E_0}+1)(e^{\beta E}+1)}(E-E_0)dE + \int_{E_0}^{\infty} \frac{e^{\beta E_0}A(E-E_0)}{(e^{\beta E_0}+1)(e^{\beta E}+1)}(E-E_0)dE$$

$$+\int_{-\infty}^{-E_0} \frac{e^{\beta E}A(E+E_0)}{(e^{-\beta E_0}+1)(e^{\beta E}+1)}(E+E_0)dE + \int_{-E_0}^{\infty} \frac{e^{-\beta E_0}A(E+E_0)}{(e^{-\beta E_0}+1)(e^{\beta E}+1)}(E+E_0)dE$$

(50)

It is easily shown that on the right hand side the first and the fourth terms cancel, as do the second and the third terms. For example, by putting $E - E_0 = x$ in the first integral we get

$$\int_{-\infty}^{0} \frac{e^{\beta(x+E_0)}A(x)}{(e^{\beta E_0}+1)(e^{\beta(x+E_0)}+1)}xdx = \int_{-\infty}^{0} \frac{e^{-\beta E_0}A(x)}{(e^{-\beta E_0}+1)(e^{-\beta(x+E_0)}+1)}xdx \xrightarrow{x \to -x} -\int_{0}^{\infty} \frac{e^{-\beta E_0}A(x)}{(e^{-\beta E_0}+1)(e^{\beta(x-E_0)}+1)}xdx$$

which is opposite in sign to what we get by using $E + E_0 = x$ in the fourth integral. It follows that $F(E_0) + F(-E_0) = 0$ and consequently

$$I_h(\phi = 0) = 0.$$  (51)

The behavior of $I_h(\phi)$ for small bias $\phi$ can be evaluated under the assumption that the dependence of the transmission probability on $\phi$ can be disregarded, i.e. $T'_{RL}(E_0, E, \phi) = T'_{LR}(E_0, E, \phi) = T'(E_0, E, \phi = 0)$. In this case the $\phi$ dependence is dominated by the Fermi function $f$. The linear term can be obtained by setting

$$f(E + e\phi) \approx f(E) - e\phi\delta(E - E_F)$$  (52)

Using this together with (51) in Eq. (29) leads to

$$I_h = -\frac{e\phi}{\pi\hbar}\int_{-\infty}^{\infty} dE_0 T'(E_0, E_F)f(E_0)(E_F - E_0) + \frac{e\phi}{\pi\hbar}\int_{-\infty}^{\infty} dE T'(E_F, E)(1 - f(E))(E - E_F)$$ (53)

Assuming again that the differential transmission probability has the form (47) we get that up to linear terms in $\phi$

$$I_h = -\frac{e\phi}{\pi\hbar}\int_{-\infty}^{E_F} dE_0 A(E_F - E_0)e^{-\beta(E_F - E_0)}f(E_0)(E_F - E_0) - \frac{e\phi}{\pi\hbar}\int_{E_F}^{\infty} dE_0 A(E_F - E_0)f(E_0)(E_F - E_0)$$

$$+\frac{e\phi}{\pi\hbar}\int_{-\infty}^{E_F} dE A(E - E_F)(1 - f(E))(E - E_F) + \frac{e\phi}{\pi\hbar}\int_{E_F}^{\infty} dE e^{-\beta(E - E_F)}A(E - E_F)(1 - f(E))(E - E_F)$$

(54)

which can be shown to vanish by the same procedure used for Eq. (50). Thus, under the assumptions made the first contribution to the Joule heating of the junction comes from an $O(\phi^2)$ term

$$
\begin{aligned}
I_h &= \frac{e^2\phi^2}{2\pi\hbar} \int_{-\infty}^{\infty} dE_0 \int_{-\infty}^{\infty} dE\, \mathcal{T}'(E_0, E)\left[f(E_0)f''(E) - f''(E_0)(1 - f(E))\right](E - E_0) \\
&= \frac{e^2\phi^2}{\pi\hbar} \int_{-\infty}^{\infty} dE_0 \int_{-\infty}^{\infty} dE\, \mathcal{T}'(E_0, E) f(E_0) f''(E)(E - E_0)
\end{aligned}
\tag{55}
$$

This contribution to the thermal current is the Joule heating. Note that for a classical diffusive resistor the factor multiplying $\phi^2$ in (55) should be the conductivity $G$, however, because some of the energy $e\phi$ is not deposited on the bridge but on the leads, this factor should be smaller than $G$ computed from $\lim_{\phi \to 0}(I/\phi)$ with $I$ given by Eq. (29). (Note that due to the presence of thermal relaxation this G is not given by Landauer formula).

**Acknowledgements**. This research was sponsored by the Wolkswagen Foundation, the U.S.-Israel Binational Science Foundation and the Israel Science Ministry. We thank Mark Ratner and Vladimiro Mujica for many discussions on thermal effects in molecular junctions. The work of DS is supported by a fellowship from the Clore Foundation. AN thanks the Institute of Theoretical Physics at UCSB for hospitality during the final stages of this work.

**Figure Captions**

Fig. 1. A classical barrier separating two particle reservoirs without (a) and with applied bias.

Fig. 2. Several possible scenarios for the potential drop profile across a molecular junction. See discussion in text.

Fig. 3. A schematic view of the model described by Eqs. (6)-(9) and the accompanying text.

Fig. 4. Different models for the potential distribution along a model molecular bridge. See text for details.

Fig. 5. The fraction $\alpha$ (Eq. (25)) of the available energy that is released as heat on the bridge, plotted against the potential bias $\phi$ for different bridge lengths N and thermal coupling parameters ($\kappa$). Full line: $N=10$, $\kappa=200 cm^{-1}$. Dashed line: $N=5$, $\kappa=200 cm^{-1}$. Dotted line: $N=10$, $\kappa=50 cm^{-1}$. Dashed-dotted line: $N=5$, $\kappa=50 cm^{-1}$. For other system parameters see text.

Fig. 6. $\phi/(1-\alpha)$ plotted against $\kappa^2$ for different bridge lengths. From bottom to top: $N=4,5,6,7,8$.

Fig. 7. Current vs. Voltage in models A (full line) and B (dashed line) of Fig. 4 for a four-site bridge. See text for the system parameters used.

Fig. 8. The electron current (up) and the heat release per second (down) computed for a system with 4 bridge units in model A (full line) and B(dashed) line. System parameters (see text) are as in Fig. 7.

Fig. 9. The heat release per transmitted electron, Eq. (30), computed for a system with 4 bridge units in model A (full line) and B(dashed line). System parameters (see text) are as in Fig. 7.

Fig. 10. The fraction $w/(e\phi)$ of available energy that is released as heat on the bridge, computed in the framework of Eqs. (28)-(30) for a system with 4 bridge units,

for models A (full line) and B (dashed line). System parameters (see text) are as in Fig. 7.

Fig. 11. The electron current (a) and the heat release per transmitted electron (b) plotted against the bridge length $N$ for $\Delta E=2000 \text{cm}^{-1}$, $\Gamma=160 \text{ cm}^{-1}$, $V=200 \text{ cm}^{-1}$, $\kappa=50 \text{ cm}^{-1}$. Full line: model A, $\phi=0.1\text{V}$; Dashed line: Model A, $\phi=0.5\text{V}$; Dash-dotted line: model B, $\phi=0.1\text{V}$; Dotted line: Model B, $\phi=0.5\text{V}$; all for $T=300\text{K}$. Line with + marks shows results for model B at $\phi=0.1\text{V}$ and $T=200\text{K}$.

Fig. 12. The fraction of heat released on the bridge per transmitted electron plotted against the thermal coupling strength $\kappa$ for different choices of models and molecular parameters. $\Delta E=2000\text{cm}^{-1}$, $V=200\text{cm}^{-1}$, $\phi=0.5\text{V}$, $T=300\text{K}$. Model A results: line with circles - $N=8$, $\Gamma=160\text{cm}^{-1}$; dotted line - $N=8$, $\Gamma=2500\text{cm}^{-1}$; line with squares - $N=4$, $\Gamma=160\text{cm}^{-1}$; dashed-dotted line - $N=4$, $\Gamma=2500\text{cm}^{-1}$. Model B results: dashed line - $N=8$, $\Gamma=160\text{cm}^{-1}$; full line $N=4$, $\Gamma=160\text{cm}^{-1}$.

Fig. 13. A schematic view of the theoretical construct used to discuss local aspect of thermal relaxation on the bridge: Each intermediate bridge level (here $k$) is coupled infinitesimally weakly to a fictitious continuous manifold K, which is used as a local energy probe.

Fig. 14. Heat released on the bridge between sites 1 and n, displayed as a function of the site index n for a system represented by model B, (see Sect. 2). Full line: the computed result for the standard bridge (see text for parameters) at $T=300\text{K}$ with a potential bias of 1eV. Dashed line: result for a system similar to the original, but with an impurity site represented by setting $E_5=E_B-0.125\text{V}$ where $E_B$ is the energy of all other bridge states in the site representation. Dotted line - same for an impurity characterized by $E_5=E_B+0.125\text{V}$.

Fig. 15. A model for analyzing temperature rise in a current carrying molecular bridge. The molecule is represented by a cylinder of length L and cross-section radius $R_2$ connecting between two surfaces (shaded areas) at $z=0$ and $z=L$ on which the temperature $T_\infty$ is given. Heat is deposited at a given rate w in on the inner cylinder of radius $R_1$. On the boundary $\rho=R_2$ either Neumann or Dirichlet boundary condition is taken according to the physical situation (see text).

Fig. 16. The temperature distribution in the cylinder representing the molecule (see text), obtained from solving Eq. (37) using $T(z=0)=T(z=L)=T_\infty$ and either $T(R_2)=T_\infty$ (dashed line) or $\left(\partial T/\partial \rho\right)_{\rho=R_2} = 0$ (full line) as boundary conditions. The other parameters used are $L=60\text{Å}$, $R_1=4\text{Å}$, $R_2=10\text{Å}$, $T_\infty=300\text{K}$ $I_h=10^{10}\text{eV/s}$ and $\sigma_h = 3.5\cdot 10^{-4}$ cal/(s·cm·K).

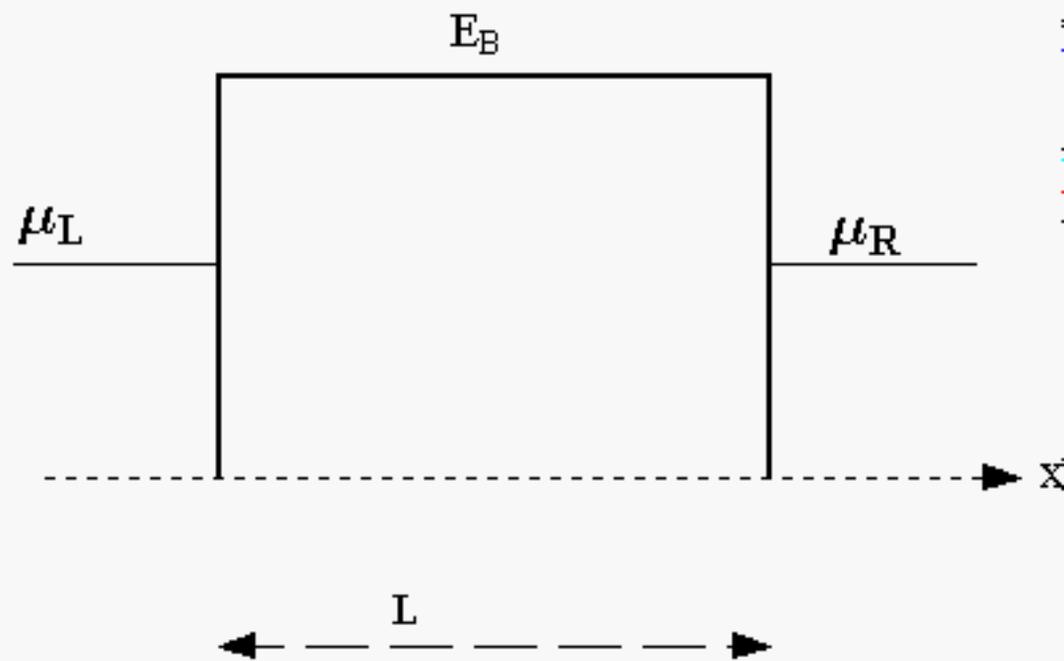

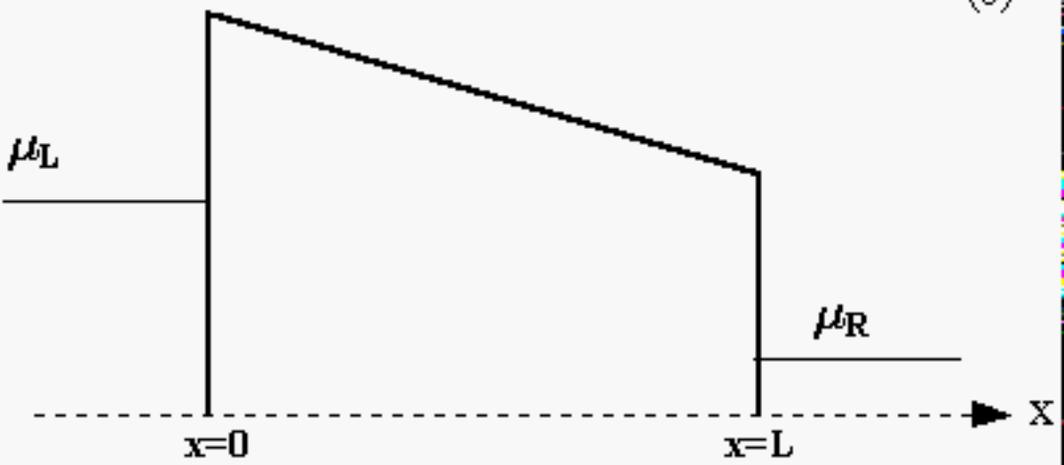

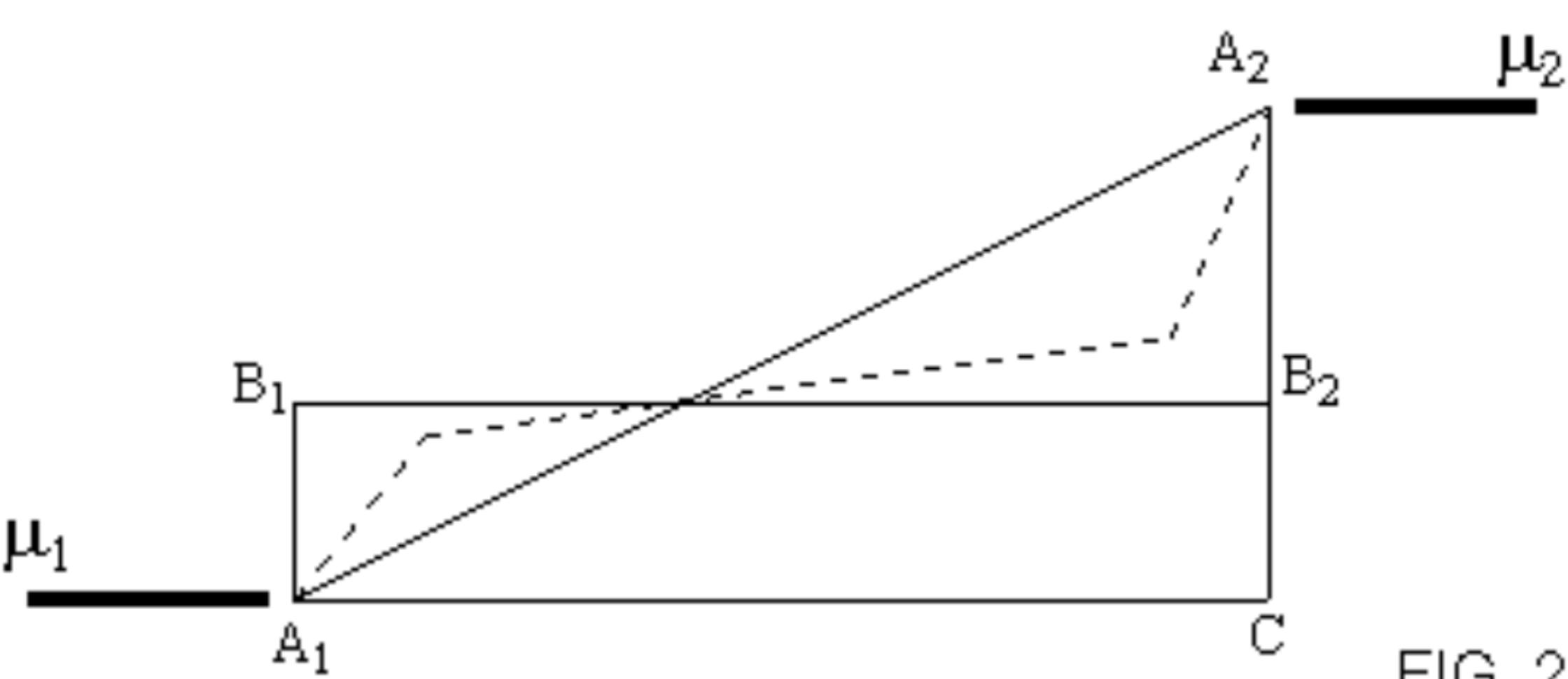

FIG. 2

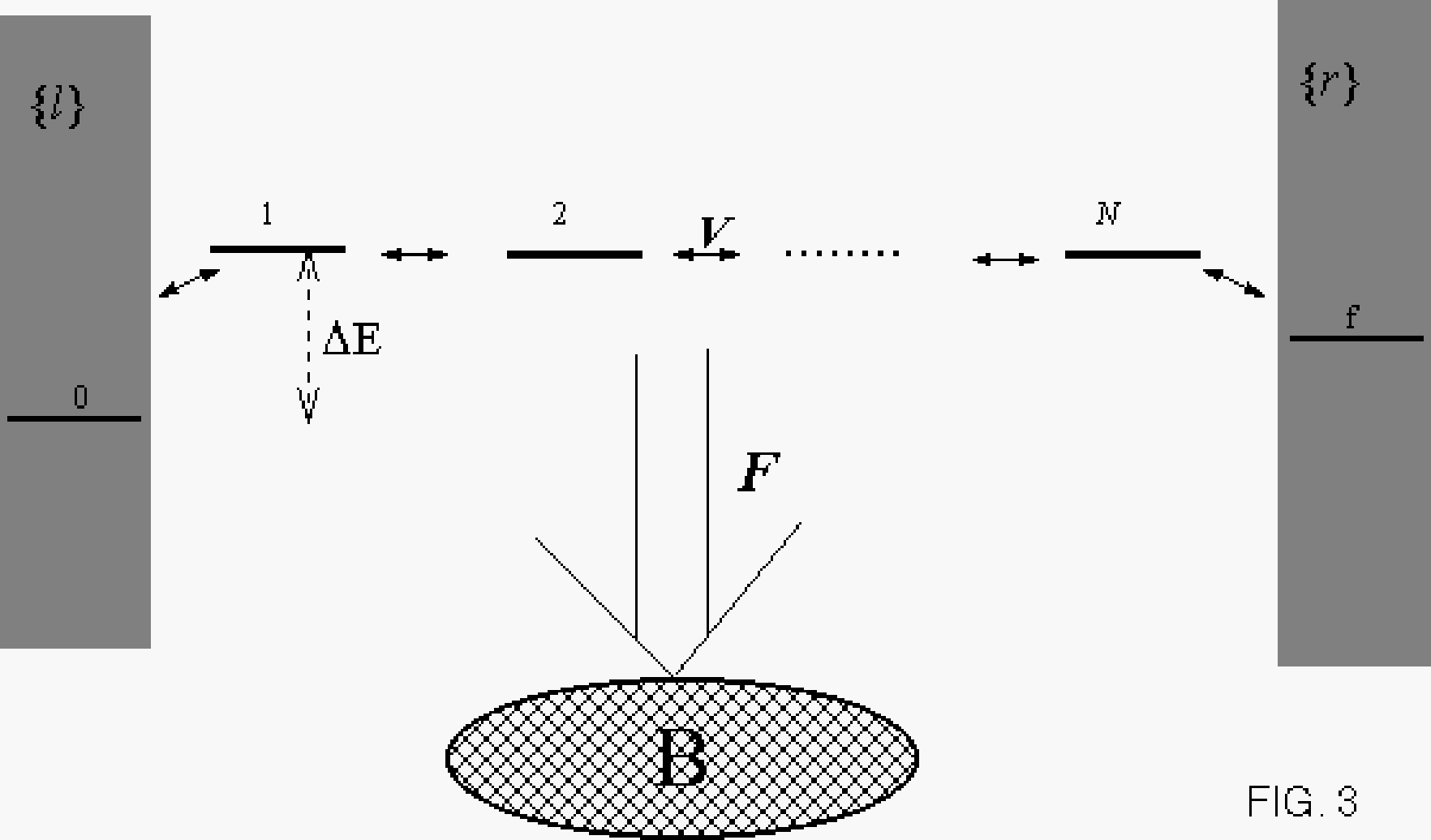

FIG. 3

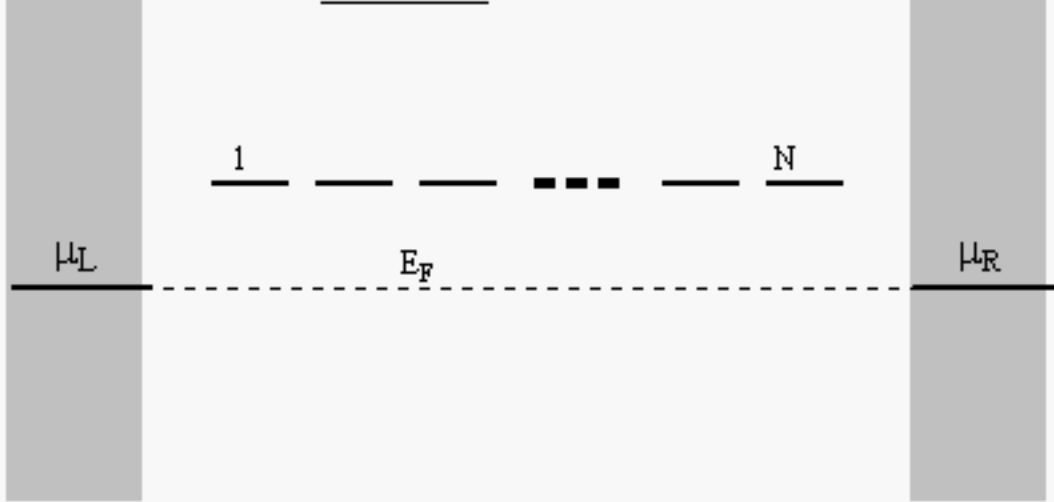
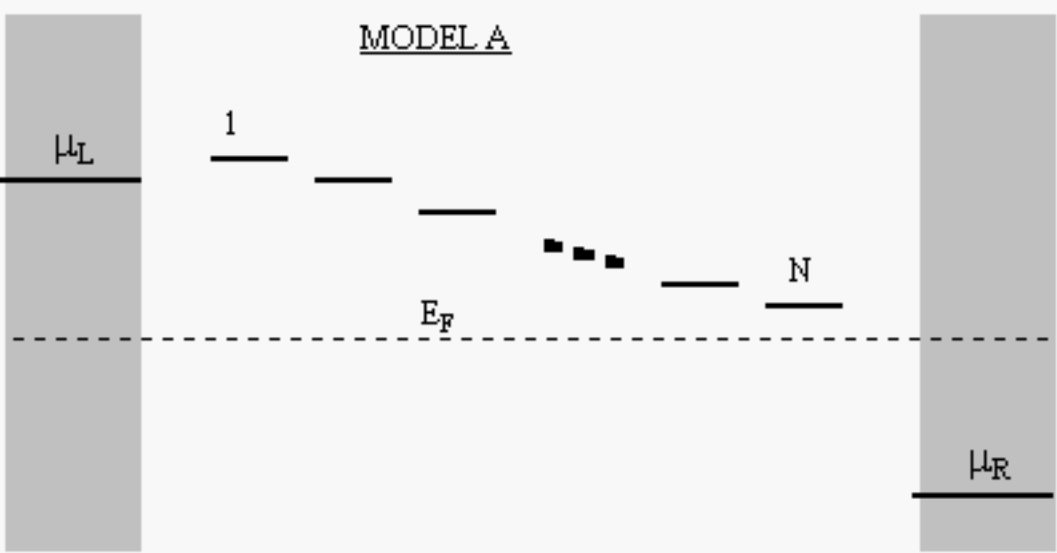
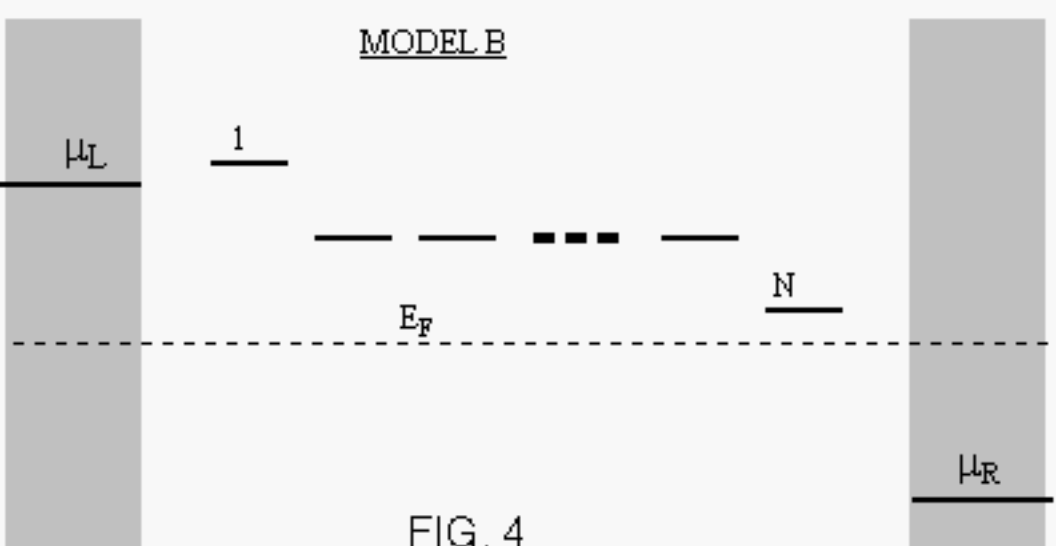

FIG. 4

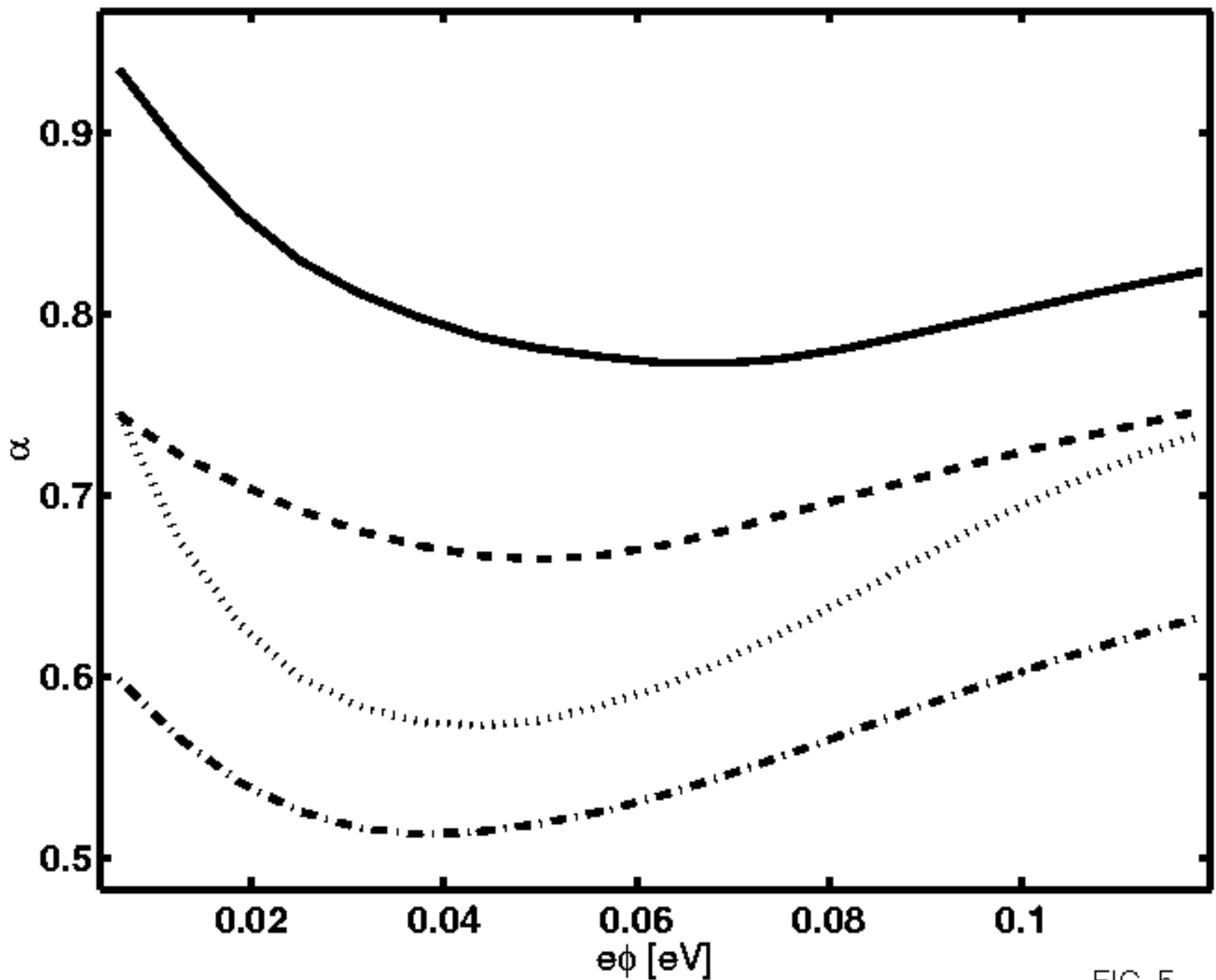

FIG. 5

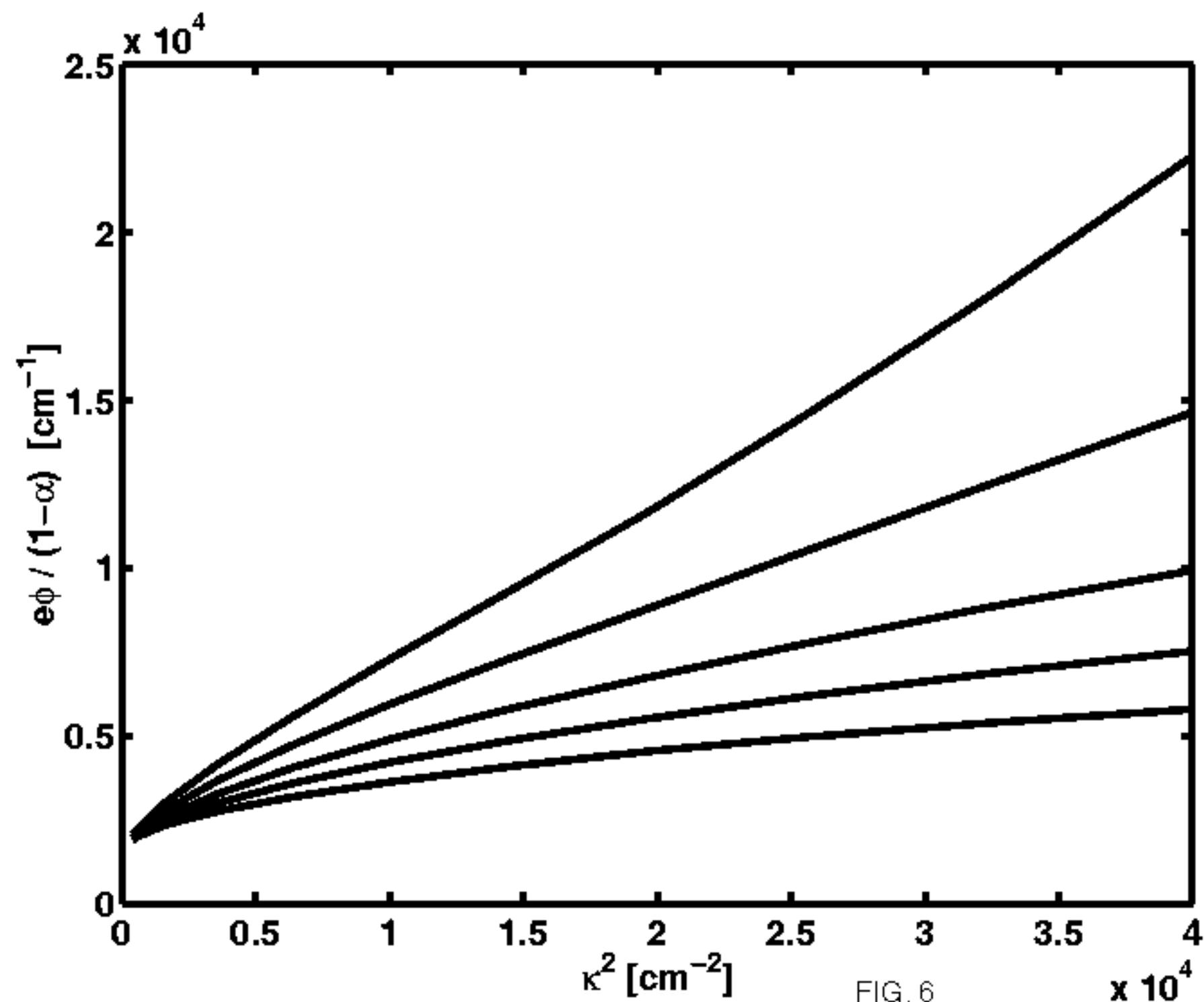

FIG. 6

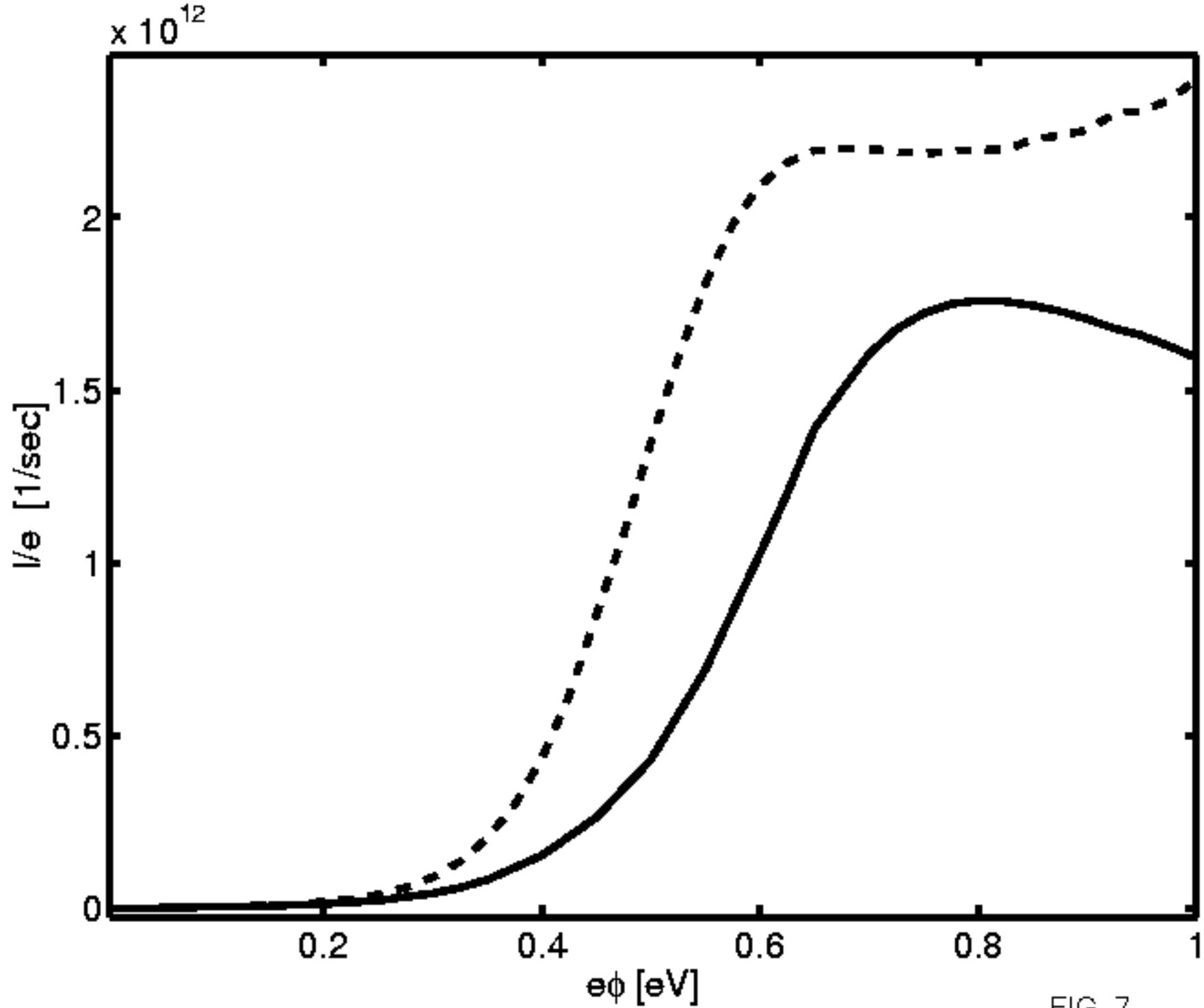

FIG. 7

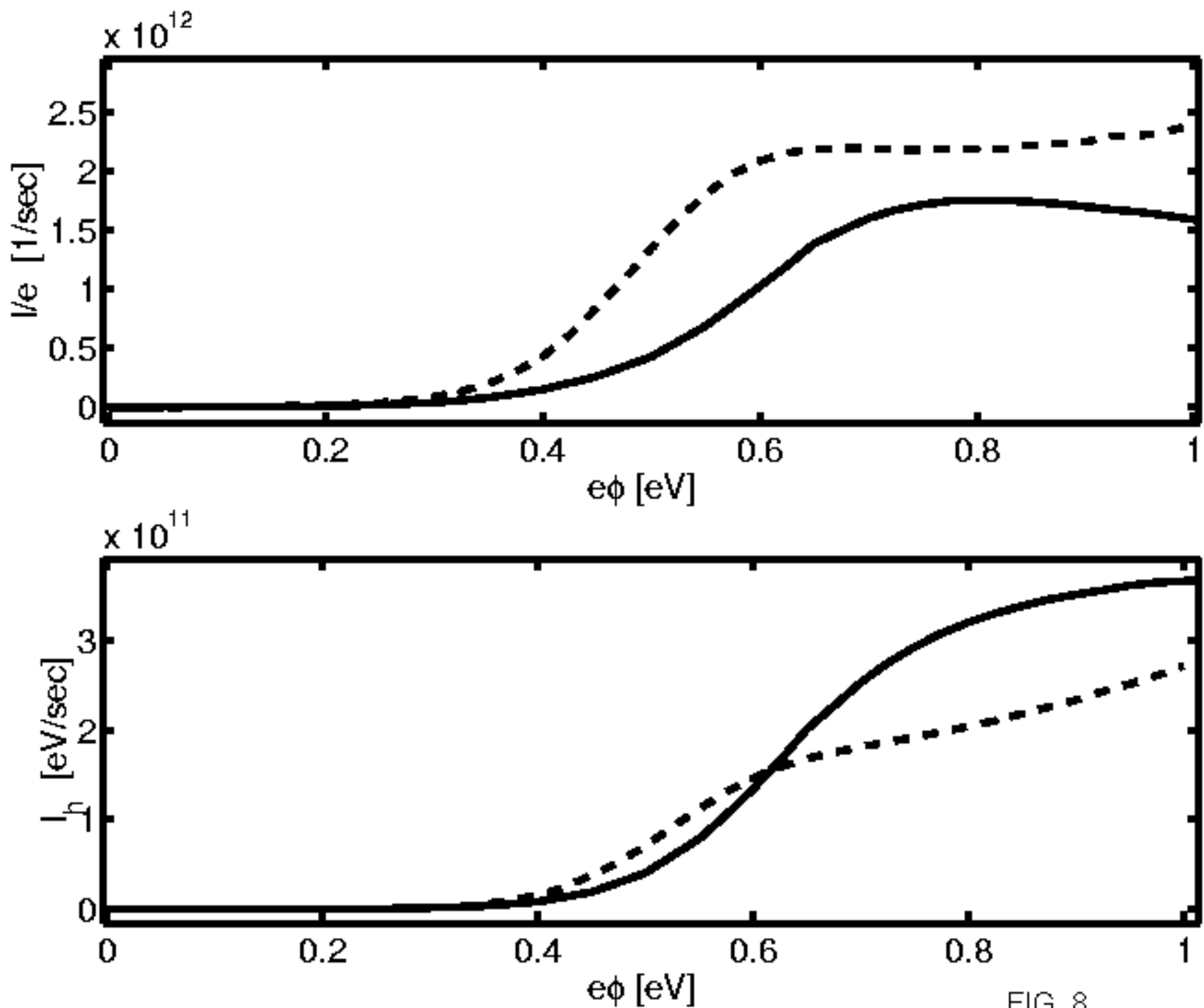

FIG. 8

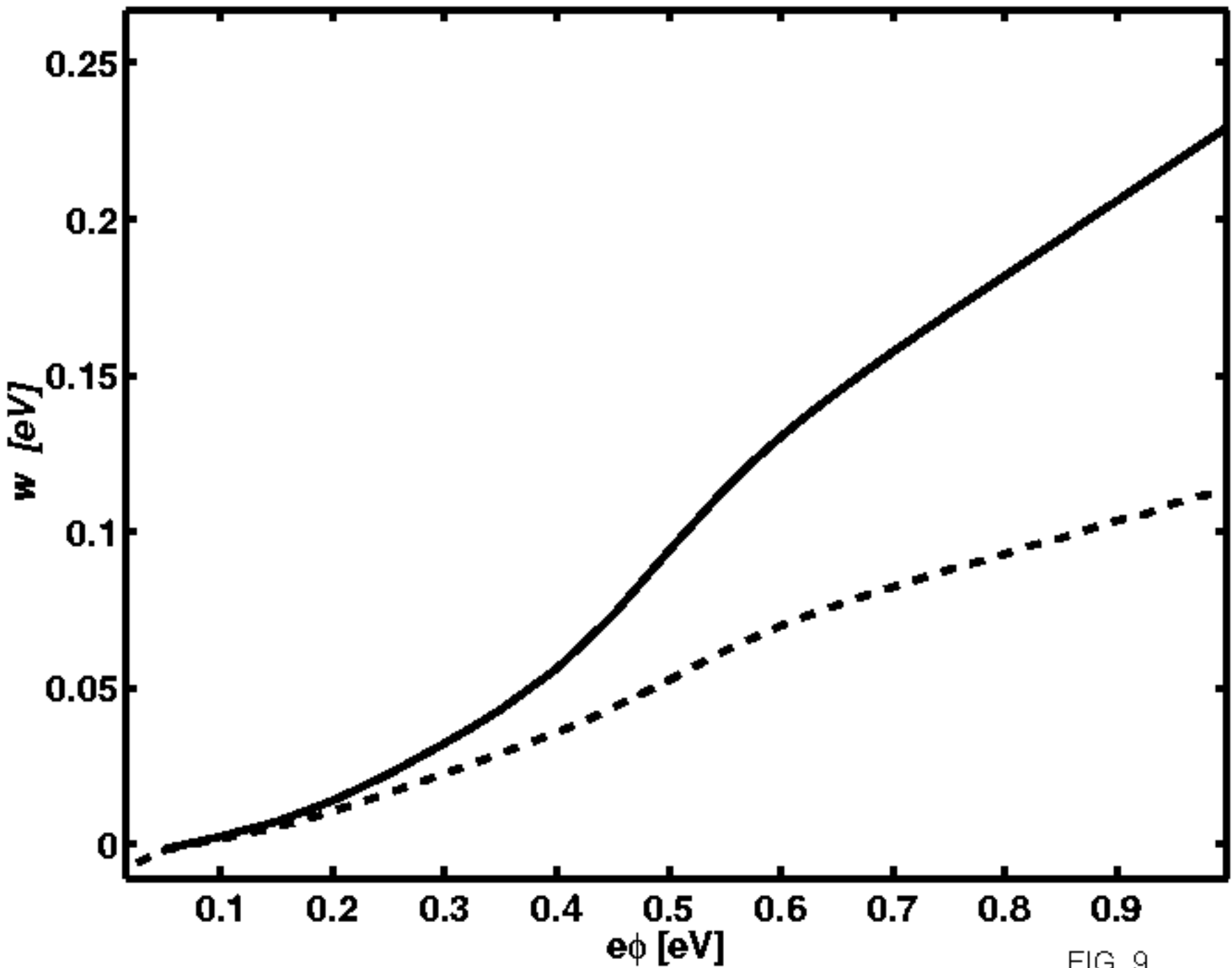

FIG. 9

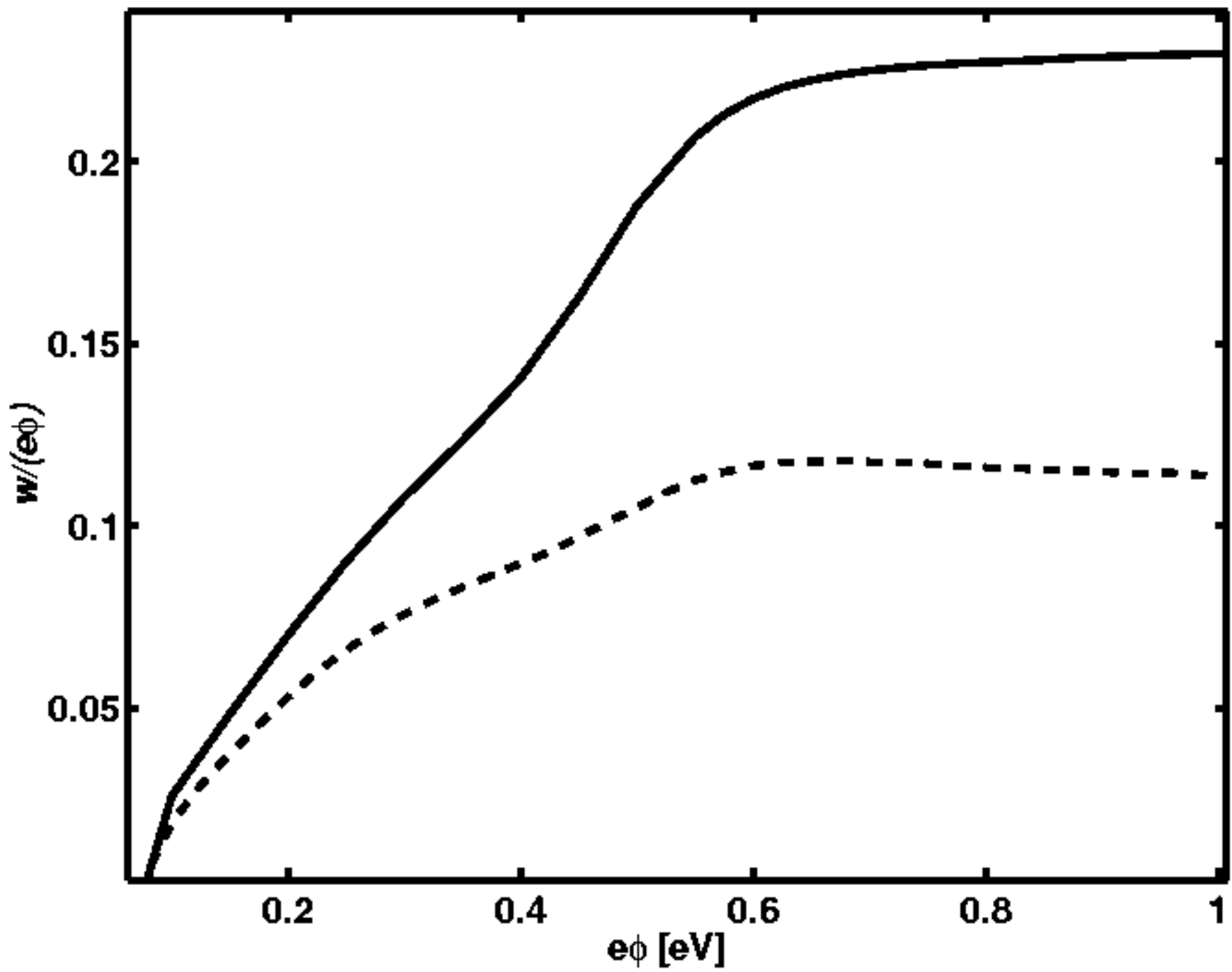

FIG. 10

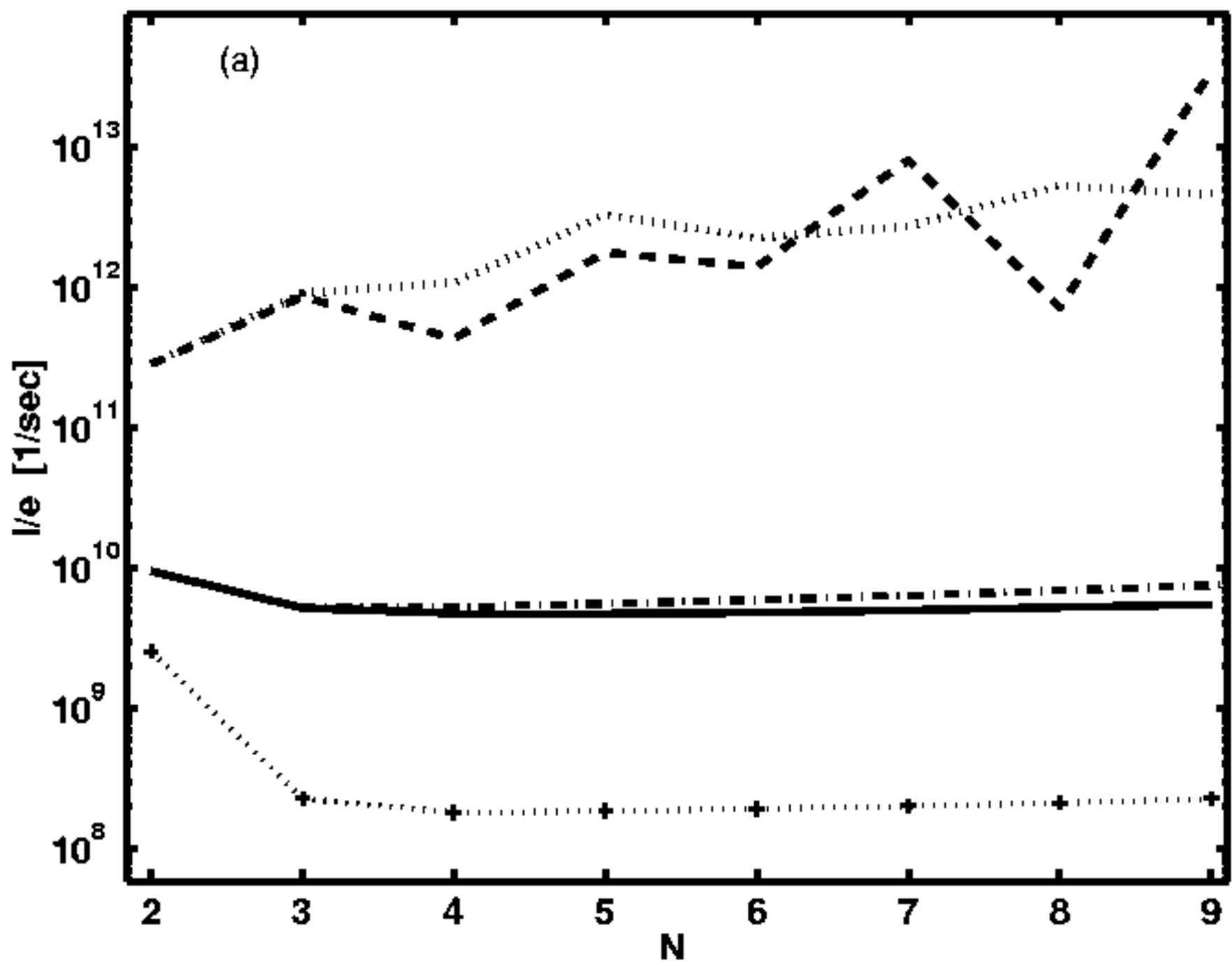

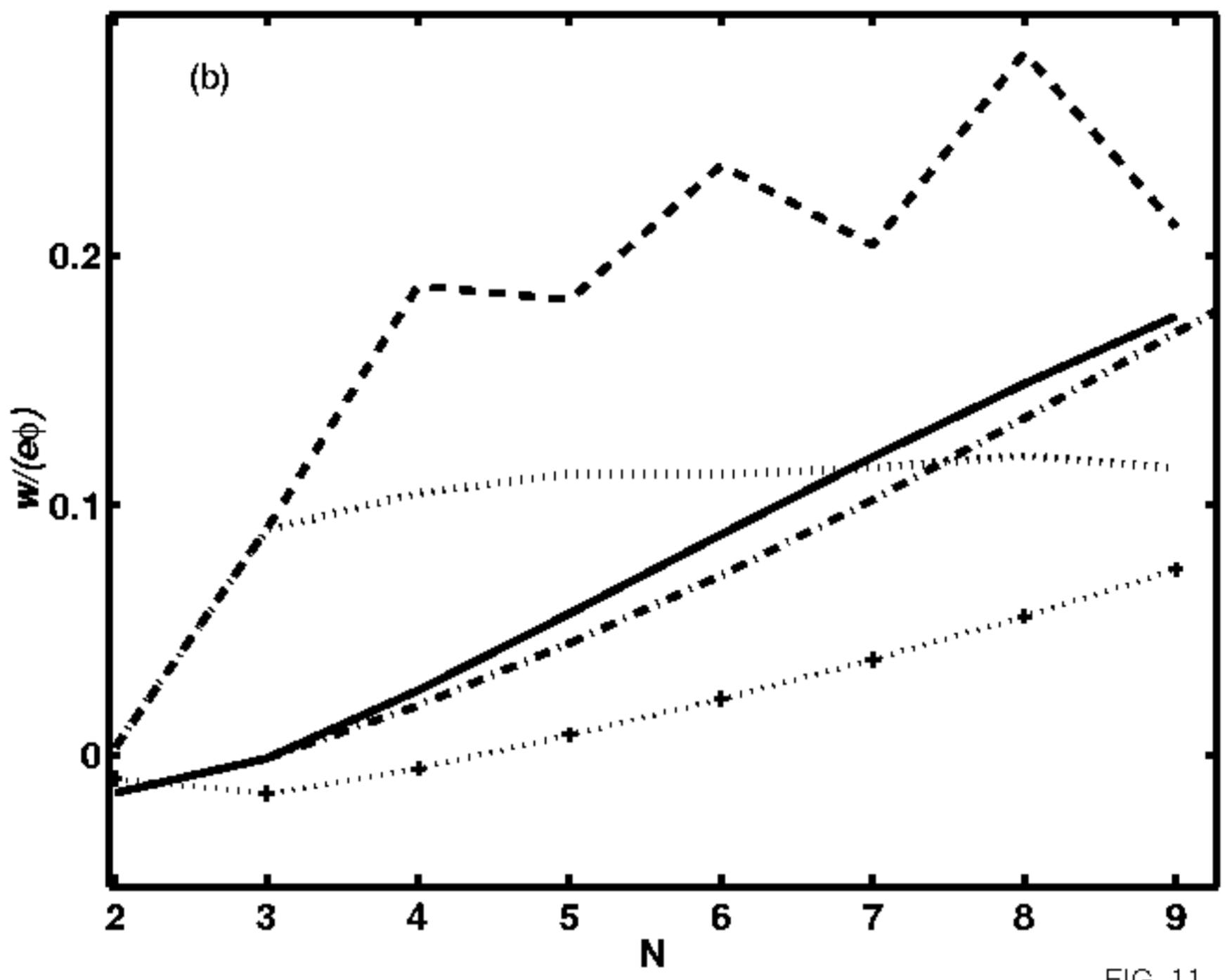

FIG. 11

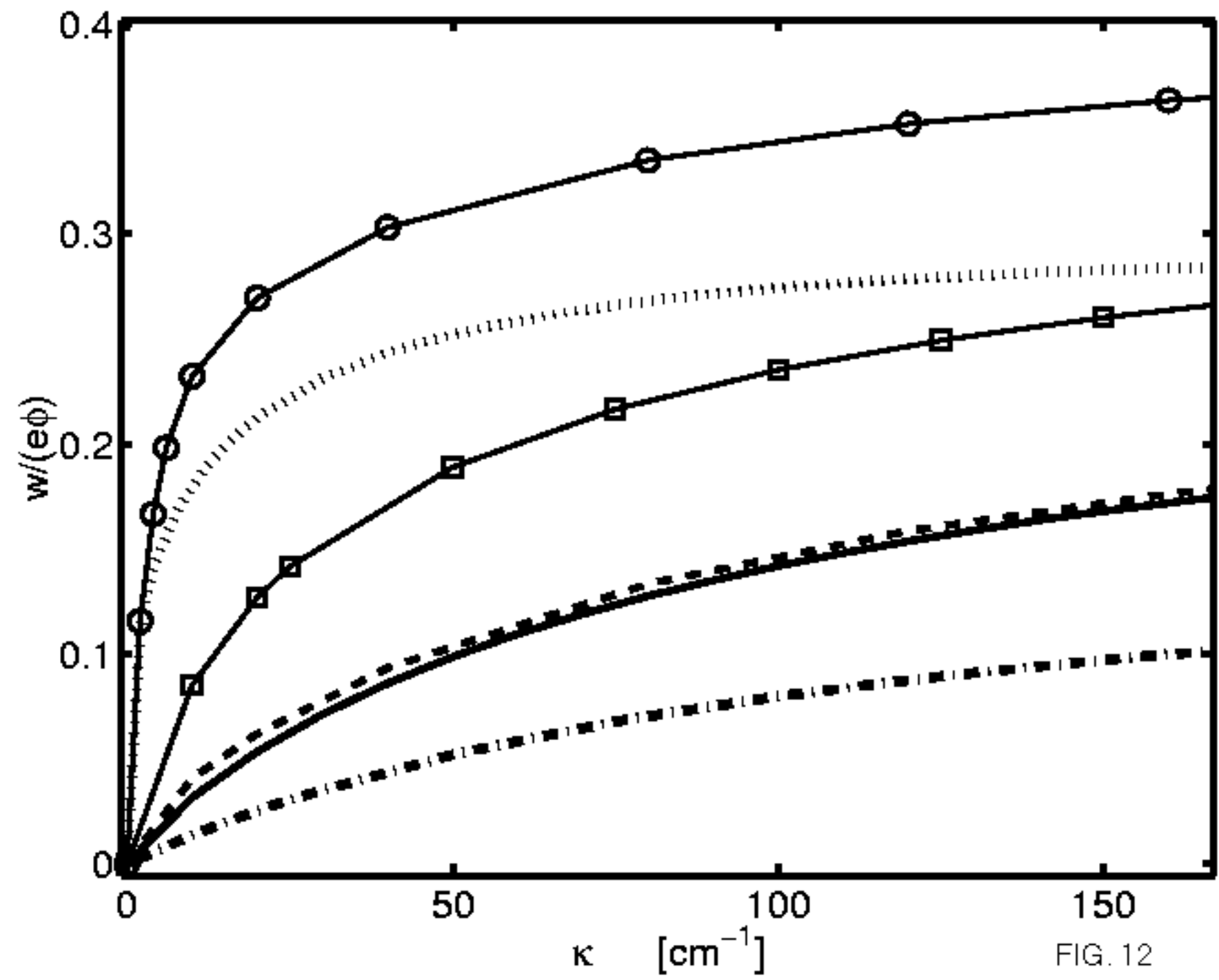

FIG. 12

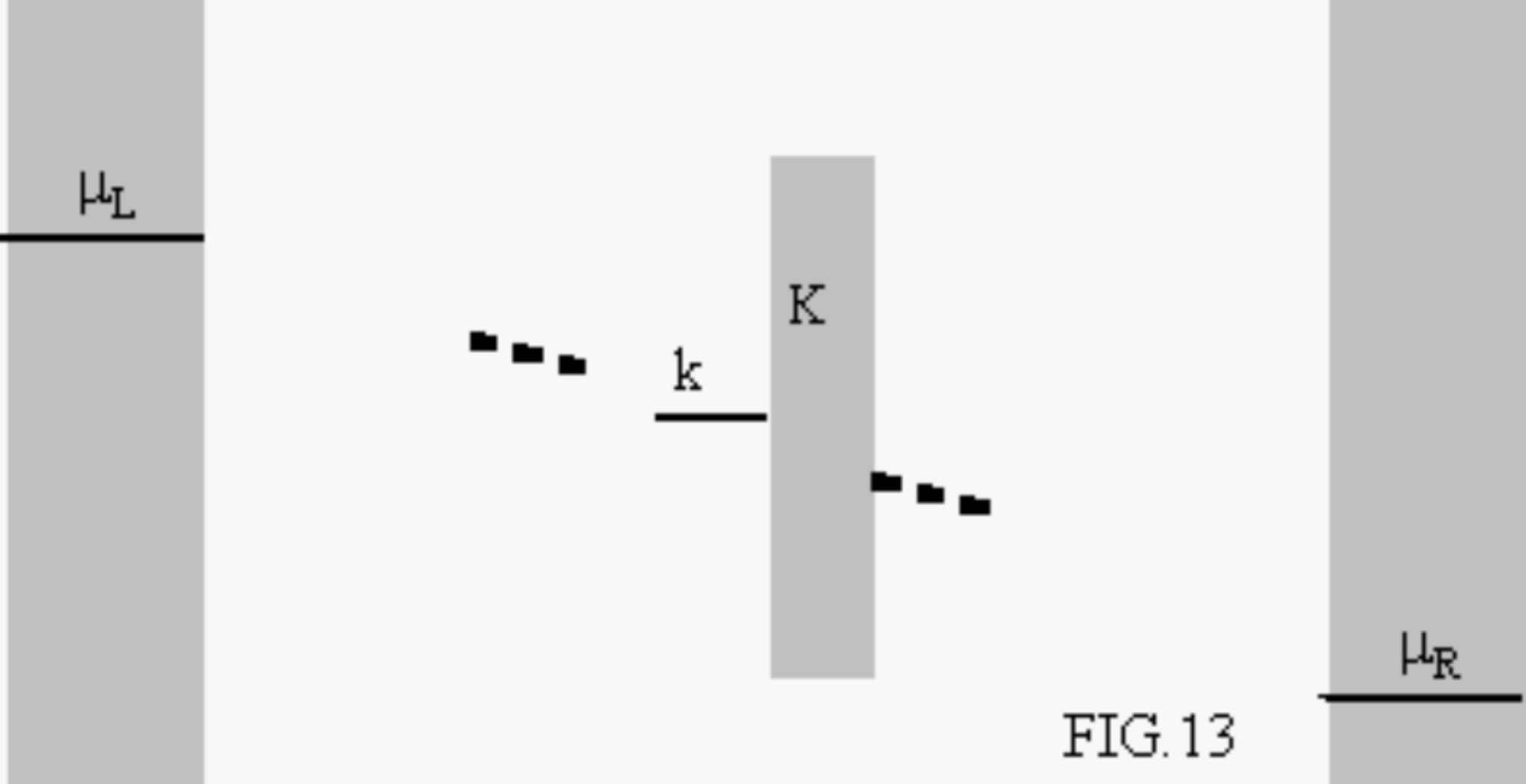

FIG. 13

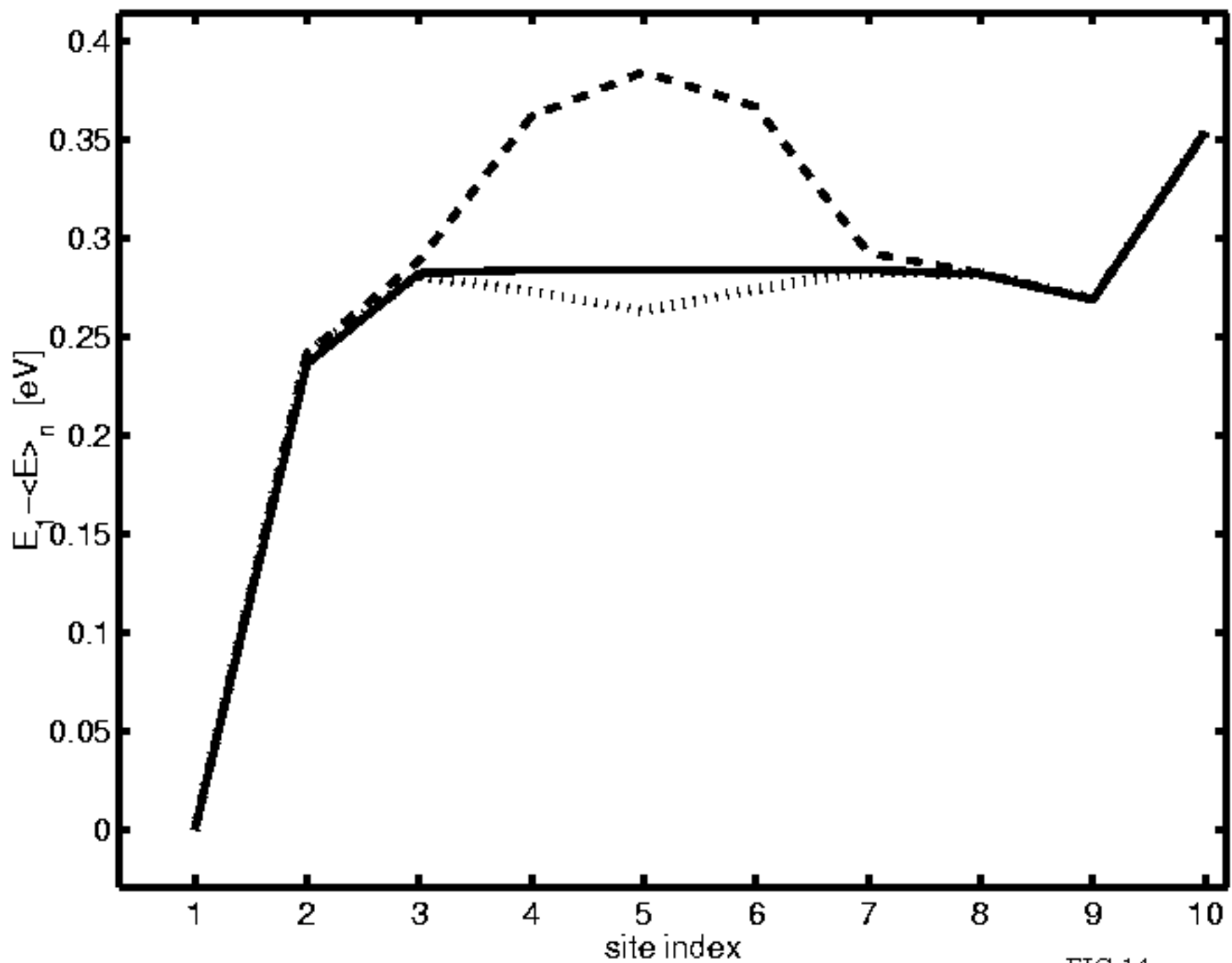

FIG.14

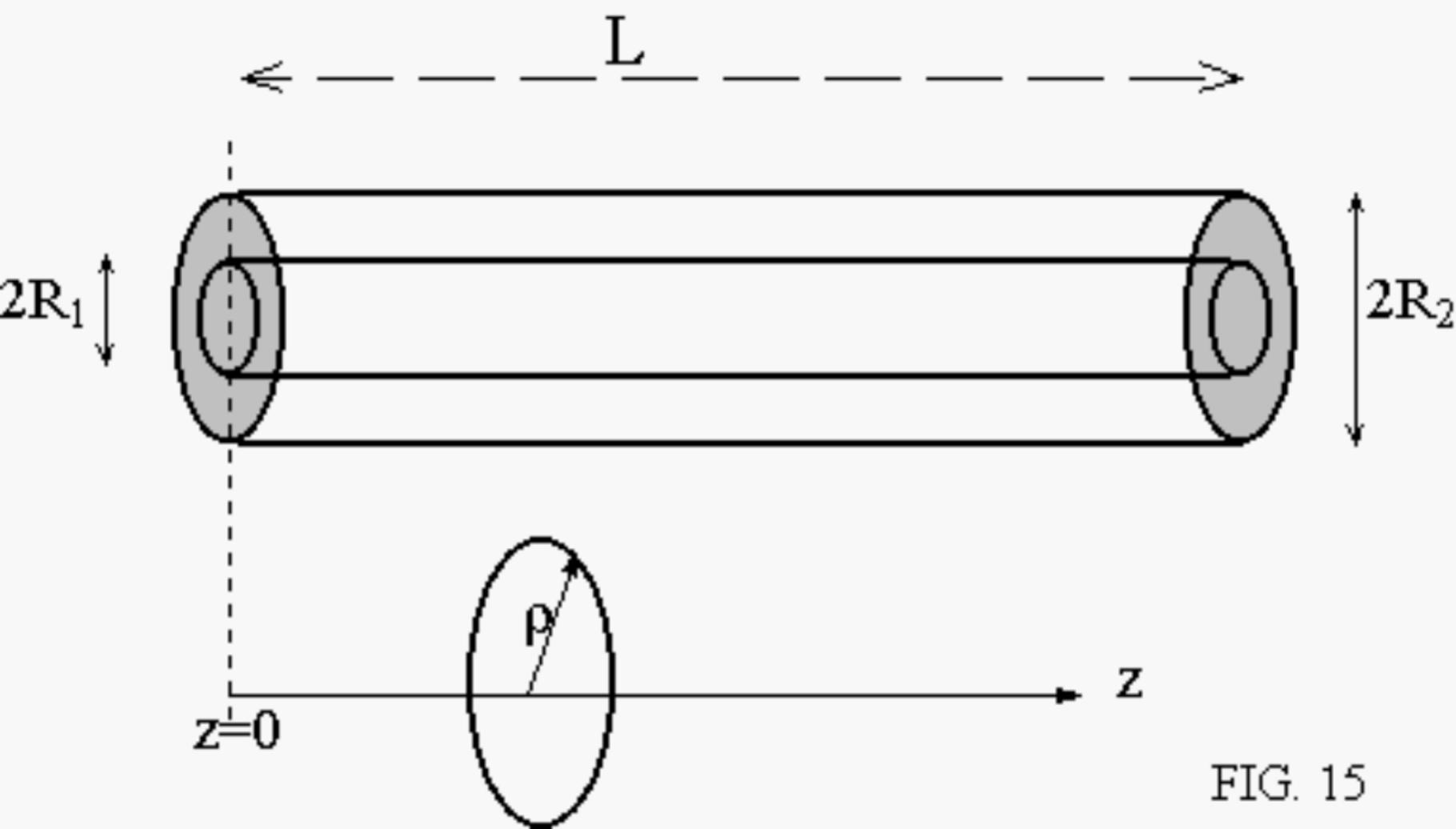

FIG. 15

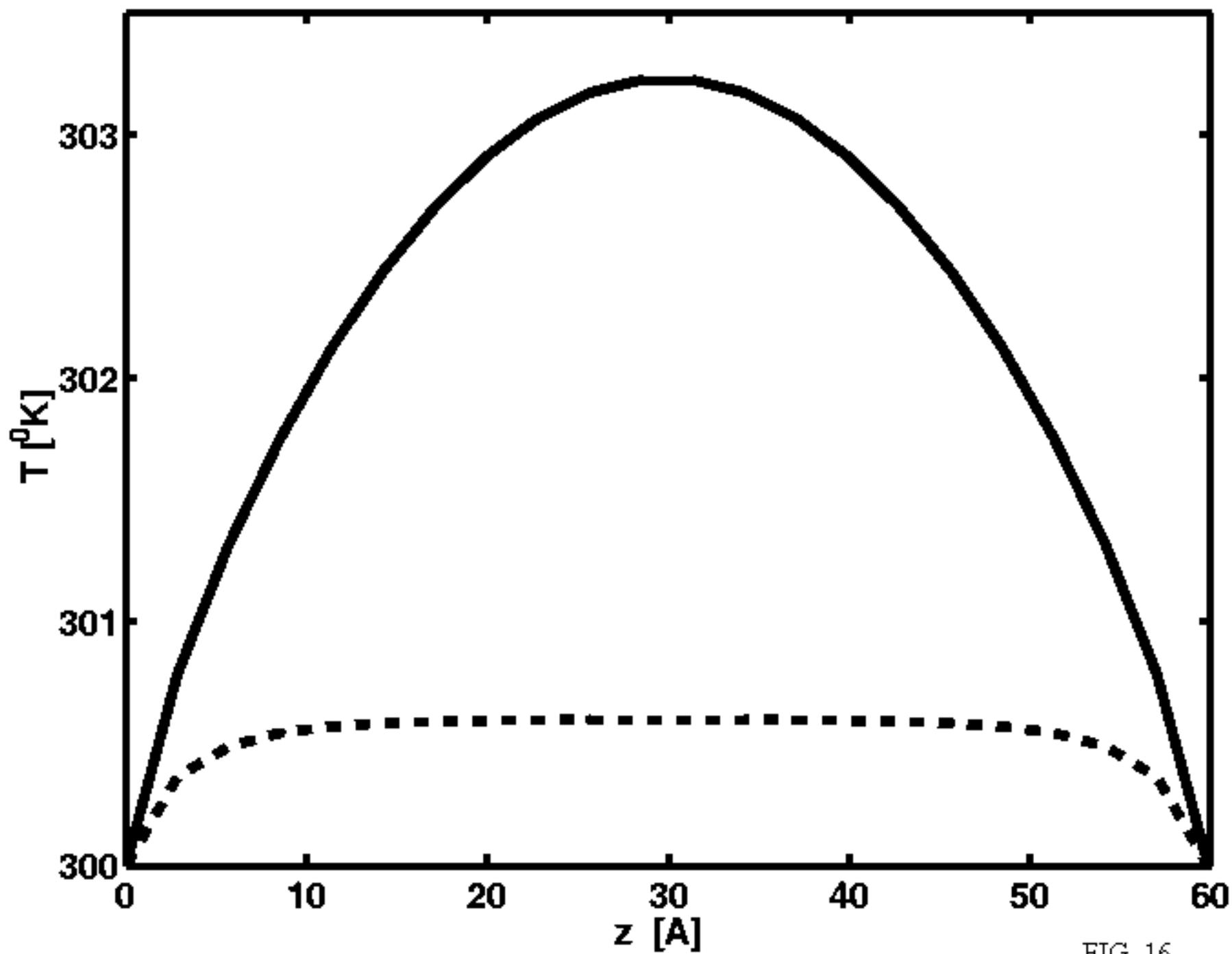

FIG. 16